\documentclass{emulateapj}

\shorttitle{Oxygen Abundances in Metal-Poor Stars}
\shortauthors{Mel\'endez, Shchukina, Vasiljeva \& Ram\'{\i}rez}

\begin{document}
\title {{\it Permitted} Oxygen Abundances and the Temperature Scale of
Metal-Poor Turn-Off Stars}

\newcommand{\teff}{T$_{\rm eff}$ }
\newcommand{\tsin}{T$_{\rm eff}$}
\newbox\grsign \setbox\grsign=\hbox{$>$} \newdimen\grdimen \grdimen=\ht\grsign
\newbox\simlessbox \newbox\simgreatbox
\setbox\simgreatbox=\hbox{\raise.5ex\hbox{$>$}\llap
     {\lower.5ex\hbox{$\sim$}}}\ht1=\grdimen\dp1=0pt
\setbox\simlessbox=\hbox{\raise.5ex\hbox{$<$}\llap
     {\lower.5ex\hbox{$\sim$}}}\ht2=\grdimen\dp2=0pt\def\simgreat{\mathrel{\copy\simgreatbox}}
\def\simless{\mathrel{\copy\simlessbox}}
\newbox\simppropto
\setbox\simppropto=\hbox{\raise.5ex\hbox{$\sim$}\llap
     {\lower.5ex\hbox{$\propto$}}}\ht2=\grdimen\dp2=0pt
\def\simpropto{\mathrel{\copy\simppropto}}

\author{Jorge Mel\'endez\altaffilmark{1}} 
\affil{Department of Astronomy, California Institute of Technology, 
MC 105--24, Pasadena, CA 91125, USA \\
Research School of Astronomy \& Astrophysics, 
Mt. Stromlo Observatory, Cotter Rd., Weston Creek, ACT 2611, Australia} 
\email{jorge@mso.anu.edu.au}

\author{Nataliya G. Shchukina, Irina E. Vasiljeva}
\affil{Main Astronomical Observatory, National Academy of Sciences, 
27 Zabolotnogo Street, Kiev 03680, Ukraine}\email{shchukin@mao.kiev.ua, vasil@mao.kiev.ua} \and

\author{Iv\'an Ram\'{\i}rez\altaffilmark{1}}
\affil{Department of Astronomy, University of Texas at Austin, 
RLM 15.306, TX 78712-1083, USA}\email{ivan@astro.as.utexas.edu}

\altaffiltext{1}{Affiliated with the Seminario Permanente 
de Astronom\'ia y Ciencias Espaciales of the 
Universidad Nacional Mayor de San Marcos, Peru}

\slugcomment{Submitted to the The Astrophysical Journal}
\slugcomment{Send proofs to:  J. Melendez}

\begin{abstract}

We use high quality VLT/UVES published data of the 
permitted  \ion{O}{1} triplet and \ion{Fe}{2} lines 
to determine oxygen and iron abundances in 
unevolved (dwarfs, turn-off, subgiants) metal-poor
halo stars.
The calculations have been performed both in LTE and NLTE,
employing effective temperatures obtained with the 
new infrared flux method (IRFM) temperature scale 
by Ram\'{\i}rez \& Mel\'endez, and surface gravities 
from {\it Hipparcos} parallaxes and theoretical isochrones. 
A new list of accurate transition probabilities for \ion{Fe}{2} lines,
tied to the absolute scale defined by laboratory measurements, 
has been used.
Interstellar absorption has been carefully taken into account by employing
reddening maps, stellar energy distributions and Str\"omgren photometry.

We find a plateau in the oxygen-to-iron ratio over 
more than two orders of magnitude in iron abundance 
($-3.2 <$ [Fe/H] $< -0.7$), with a mean
[O/Fe] = 0.5 dex ($\sigma$ = 0.1 dex), independent of
metallicity, temperature and surface gravity.
The flat [O/Fe] ratio is mainly due to the use of adequate NLTE corrections
and the new IRFM temperature scale, which, for metal-poor F/early G dwarfs
is hotter than most \teff scales used in previous studies 
of the \ion{O}{1} triplet. 

According to the new IRFM \teff scale, the temperatures of
turn-off halo stars strongly depend on metallicity,
a result that is in excellent qualitative and quantitative
agreement with stellar evolution calculations,
which predict that the \teff of the turn-off at [Fe/H] = $-3$ is about 
600-700 K higher than that at [Fe/H] = $-1$. 
Recent determinations of H$\alpha$ temperatures 
in turn-off stars are in excellent relative
agreeement with the new IRFM \teff scale 
in the metallicity range $-2.7 <$ [Fe/H] $< -1$,
with a zero point difference of only 61 K.

\end{abstract}


\section{Introduction}

The importance of the oxygen abundance in metal-poor
stars has been strongly emphasized in the literature,
as well as the problems related to its determination. 
There is currently no consensus as to whether 
the [O/Fe] ratio in halo stars is 
approximately constant ([O/Fe] $\approx$ 0.4 - 0.5)
or steeply increases with decreasing [Fe/H] 
([O/Fe]$\approx$ -0.35 [Fe/H]). The discrepancy is due 
to problems in the modeling of the different spectral features
used to estimate the oxygen abundance, which are
very hard to detect in the same star.

Due to its high excitation potential,
the permitted \ion{O}{1} triplet at 0.77$\mu$m is mainly observed in FG dwarfs and subgiants, 
being very faint (yet detectable) in metal-poor early K stars. 
The low excitation potential forbidden lines [\ion{O}{1}] at 0.63 $\mu$m
are detected in giants and cool subgiants,
and barely detectable in dwarfs. 
Molecular OH lines are observed in the 
ultraviolet (electronic transitions at 0.3 $\mu$m) and in the infrared 
(vibrational-rotational lines at 1.5 and 3 $\mu$m). 
Both observations are difficult: the UV OH lines
are close to the atmospheric cutoff and most CCDs have low
sensitivity in the UV, while the weak IR OH lines require extremely high S/N.
The UV OH lines are relatively strong and observed in FGK
metal-poor stars, while the very weak IR OH lines 
are only observed in cool stars with \teff $<$ 5000 K.

The observed spectra by themselves do not allow a direct
measurement of the oxygen abundance, so a careful modeling
must be performed. Since the spectral
features are sensitive to both effective temperature and
surface gravity, reliable atmospheric parameters must be used.
The analysis is complicated by the presence 
of both NLTE (mainly affecting the \ion{O}{1} triplet) 
and granulation (crucial for the molecular OH lines) effects.

Analyses of the forbidden [\ion{O}{1}] 
(e.g. Barbuy 1988; Sneden et al. 1991) and 
infrared OH lines (Mel\'endez, Barbuy \& Spite 2001)
seem to show a flat [O/Fe], or a small increase 
of about 0.1-0.2 dex per a factor of 10 (1 dex) in metallicity,
from [O/Fe]$\approx$ 0.3 at [Fe/H] = $-$1.5 to [O/Fe] $\approx$ 0.4-0.5 
at [Fe/H] = $-$2.5 (Sneden \& Primas 2001; Mel\'endez \& Barbuy 2002).
Yet, the 1D analysis of the forbidden line by Nissen et al. (2002) 
shows a continuous increase of [O/Fe] for lower metallicities, 
reaching [O/Fe] $\approx$ +0.7 at [Fe/H] = $-2.5$, 
but after correcting for 3D effects they found 
a plateau at [O/Fe] $\approx$ +0.3 in the 
range $-$2.0 $<$ [Fe/H] $<$ $-$0.7,
and an increase in [O/Fe] for lower [Fe/H]. 
The recent 1D analysis of [\ion{O}{1}] in subgiants by
Garc\'{\i}a P\'erez et al. (2005) shows a mean [O/Fe] $\approx$ +0.5,
with a very small increase of 0.09 dex in [O/Fe] for a decrease of 1 dex in [Fe/H].

On the other hand, studies of the permitted \ion{O}{1} triplet 
(e.g. Abia \& Rebolo 1989; Cavallo, Pilachowski \& Rebolo 1997;
Israelian et al. 1998, 2001; Mishenina et al. 2000) 
and the ultraviolet OH lines (Israelian et al. 1998, 2001; Boesgaard et al. 1999)
show a steep monotonic increase of [O/Fe] for lower metallicities, 
reaching [O/Fe] = +1.1 at [Fe/H] = $-3$. However, other
studies of the triplet show a lower (or zero) slope. 
For example, the analysis of the triplet in turn-off stars
by Akerman et al. (2004, hereafter Ake04) shows only a 
mild increase of [O/Fe] with metallicity,
with [O/Fe] $\approx$ +0.4 at [Fe/H] = $-$1 and 
[O/Fe]$\approx$ +0.7 at [Fe/H] = $-$3.

Asplund \& Garc\'{\i}a P\'erez (2001) have shown 
that 1D model atmospheres overestimate the abundances obtained
from UV OH lines by as much as 0.9 dex at [Fe/H] = $-3$ (see also
$\S$8.2 and Asplund 2005), and when the UV OH lines are 
analyzed with 3D model atmospheres,
the [O/Fe] ratio in metal-poor dwarfs is in good
agreement with the results obtained in giants from the
forbidden lines.
Note also that the 1D analysis of the UV OH lines
by Bessell et al. (1991) resulted in low [O/Fe] ratios,
as is also the case of the work on 
UV OH lines in subgiant stars by Garc\'{\i}a P\'erez et al. (2005),
which shows 1D oxygen-to-iron ratios of [O/Fe] $\approx$ +0.5 dex.
Hence, analyses of three oxygen abundance 
features (UV OH lines, IR OH lines, [\ion{O}{1}] lines) roughly agree 
in a somewhat flat (and low) [O/Fe] ratio in the
metallicity range $-3 <$ [Fe/H] $< -$1. However, the high
oxygen abundances obtained from the \ion{O}{1} triplet have been
difficult to reconcile with the low oxygen abundance
derived from the forbidden lines (Spite \& Spite 1991).
Nissen et al. (2002) found that 1D analysis results in
agreement between the [\ion{O}{1}] line and the triplet,
but when the lines are analyzed employing 3D model atmospheres
the oxygen abundances obtained from the triplet 
are about 0.2 dex larger than those from the [\ion{O}{1}] line
(Nissen et al. 2002). 

The disagreement between the \ion{O}{1} triplet
and the forbidden [\ion{O}{1}] lines also occurs in
metal-poor giants and subgiants 
(Cavallo et al. 1997; Fulbright \& Johnson 2003; Garc\'{\i}a P\'erez et al. 2005).
Israelian et al. (2004) undertook the first NLTE analysis
in very metal-poor CN-rich giants, showing that
there are serious problems with standard 1D model atmospheres
of those stars, even when NLTE effects are taken into account,
producing a serious conflict between the oxygen abundances 
obtained from the forbidden line and the 
triplet (Israelian et al. 2004). 
The use of CN-enhanced model atmospheres
have an important impact on the thermal 
structure of model atmospheres of metal-poor
CN-enhanced ([C/Fe] = +2, [N/Fe] = +2) giants 
(Masseron et al. 2005), which affect the abundances from a
few tenths of a dex up to 1.5 dex in extreme cases,
with respect to standard (solar-scaled abundances)
1D models (Masseron et al. 2005). Hence, for metal-poor cool giants with extreme
compositions, model atmospheres computed with
the corresponding CNO overabundances may partly relieve the
problems found by Israelian et al. (2004).

The use of 3D + NLTE seems to improve the situation 
for the oxygen abundances derived from the \ion{O}{1} triplet
and the [\ion{O}{1}] line. Employing 1.5D NLTE calculations
in a 3D model atmosphere of the metal-poor star HD 140283, 
Shchukina, Trujillo Bueno \& Asplund (2005) have found that 
the oxygen abundances obtained from the 
\ion{O}{1} and [\ion{O}{1}] lines agree within $\approx$ 0.1 dex,
which is an important achievement considering the 
observational uncertainties for the [\ion{O}{1}] line.

Previous studies of the \ion{O}{1} triplet employing
a hot temperature scale resulted in a flat and low
[O/Fe] (King 1993; Carretta, Gratton \& Sneden 2000).
Note that another reason why Carretta et al. (2000) 
obtained low [O/Fe] ratios is the use
of large NLTE corrections of the \ion{O}{1} triplet by
Gratton et al. (1999), as pointed out by Takeda (2003).

King (2000) studied the influence of stellar parameters 
and NLTE iron abundances (Th\'evenin \& Idiart 1999) 
on the [O/Fe] ratio obtained from UV OH, [\ion{O}{1}]
and \ion{O}{1} lines. The [O/Fe] values obtained by King (2000)
using the \ion{O}{1} triplet and the UV OH lines are
considerably reduced with respect to the high [O/Fe] ratios
of the original studies (Tomkin et al. 1992; Boesgaard et al. 1999),
but they are still about 0.1-0.2 dex larger than those obtained 
from the [\ion{O}{1}] lines (King 2000). Partly the low [O/Fe] ratios
obtained by King (2000) can be explained by the large 
NLTE \ion{Fe}{1} corrections by Th\'evenin \& Idiart (1999), 
which increase the iron abundance by +0.2-0.3 dex, hence
lowering [O/Fe] by 0.2-0.3 dex.

The preliminary NLTE analysis of the \ion{O}{1} triplet
by Primas et al. (2001) also resulted in a 
flat [O/Fe] ratio for halo dwarfs down to [Fe/H] = $-$2.4,
independent of the \teff scale employed. Primas et al.
found [O/Fe] $\approx$ +0.4 using the \teff scale by
Alonso et al. (1996), and [O/Fe] $\approx$ +0.5 with 
both the \teff scale by Carney (1983) and 
temperatures from Th\'evenin \& Idiart (1999). 

The observed [O/Fe] ratio in halo stars provides tight
constraints on models of Galaxy formation 
(e.g. Tinsley 1979; Wheeler, Sneden \& Truran 1989; McWilliam 1997;
Chiappini, Romano \& Matteucci 2003). 
If the formation of the
halo was fast, then the [O/Fe] ratio of halo stars
should be roughly constant, because Type Ia supernovae,
which originate from long lived low mass stars, 
would not have had enough time to lower the [O/Fe] ratio of the ISM
from which the halo stars we observe today were formed.
Furthermore, the scatter of the [O/Fe] ratio tells us
the efficiency of the ISM in homogenizing (mixing) the 
ejecta of Type II supernovae (Scalo \& Elmegreen 2004). 

The oxygen abundance is extremely important in studies
of the production of LiBeB by Galactic cosmic ray (GCR)
spallation (e.g. Prantzos, Cass\'e \& Vangioni-Flam 1993;
Fields \& Olive 1999; Ramaty et al. 2000;
Vangioni-Flam, Cass\'e \& Audouze 2000),
especially for stars with [Fe/H] $< -2$, where current
models struggle to explain the non-zero isotopic 
$^6$Li/$^7$Li ratios recently found in metal-poor dwarfs down
to [Fe/H] = $-2.7$ (Asplund et al. 2005). The GCR production of 
$^6$Li depends on the adopted [O/Fe] ratio, and only models with 
extremely high oxygen abundances can account for the $^6$Li detection 
at [Fe/H] $= -2.7$, but introduce the problem of large
overproduction of $^6$Li at higher metallicities (Rollinde et al. 2005). 
Given the problems faced by GCR models to explain the
$^6$Li observed in halo stars, several authors have suggested 
that the origin of $^6$Li may be pre-Galactic 
(Jedamzik 2000, 2004; Asplund et al. 2005; Rollinde et al. 2005). 
A low [O/Fe] ratio in metal-poor stars brings support 
for pre-Galactic $^6$Li, since in this case GCR can not
produce enough $^6$Li (see also Prantzos 2005).

Here, we analyze recent high quality VLT/UVES 
data of the \ion{O}{1} triplet and \ion{Fe}{2} lines from
Ake04 and  Nissen et al. (2004, hereafter N04), 
respectively. We show that the use of NLTE corrections and 
the new infrared flux method (IRFM)
temperature scale by Ram\'{\i}rez \& Mel\'endez (2005a,b; hereafter RM05a,b), 
which for metal-poor turn-off stars is hotter than previous temperature scales,
results in a low and flat [O/Fe] ratio in metal-poor stars.

We also compare the IRFM temperature of turn-off stars 
(employing the new IRFM \teff scale) with those expected from stellar evolution,
showing that the strong metallicity dependence of
the turn-off temperatures are very well reproduced by 
stellar evolution models.

\section{Sample}

Our sample stars consists of 31 F and G stars, which were selected by
Ake04 and N04 to be close to the main-sequence turn-off 
and with halo kinematics. 
The equivalent widths of the \ion{O}{1} triplet and \ion{Fe}{2} lines were taken
from Ake04 and N04, respectively.
The superb VLT/UVES data were obtained at a resolving power $R$ = 60 000
with 4 pixels per spectral resolution element and 
a typical S/N (per pixel) = 200-300 (Ake04, N04).

We have verified (within the uncertainties) that the sample stars 
have halo kinematics. We computed UVW velocities from their proper motions, 
radial velocities and distances (mainly from Hipparcos parallaxes, but also
from Str\"omgren photometry and isochrones), with data obtained from the
SIMBAD database. The main uncertainty is due to the 
uncertainty in distance, which in some cases leads to an 
error of a few tens of km s$^{-1}$ in the UVW velocities.

\section{Atomic Data and Solar Abundances}

\subsection{The \ion{O}{1} triplet}
The oscillator strengths of the \ion{O}{1} triplet have been
adopted from the NIST database (Wiese, Fuhr \& Deters 1996). 
We employed an interaction broadening constant $C_6$[\ion{O}{1}] = 0.84 $\times 10^{-31}$,
obtained from the collision broadening cross sections 
given by Barklem, Piskunov \& O'Mara (2000). Our \ion{O}{1} atomic model 
for the NLTE calculations is based on data by
Carlsson \& Judge (1993). The NLTE spectrum synthesis code
NATAJA was employed in the present study. For details see
Shchukina et al. (2003, 2005) and Shchukina \& Trujillo Bueno (2001).

Equivalent widths ($W_\lambda$) of the \ion{O}{1} triplet have
been measured employing the National Solar Observatory (NSO) 
FTS solar flux spectrum by Hinkle et al. (2000), which is 
essentially the same spectrum previously published by
Kurucz et al. (1984, hereafter K84), but corrected for 
telluric absorption at $\lambda >$ 5000 \AA. 
We have checked that the NSO FTS data are in the same 
scale as those of the VLT/UVES data, using a high 
resolution (R $\approx$ 10$^5$) UVES reflected solar spectrum.\footnote{
http://www.eso.org/observing/dfo/quality/UVES/}
As shown in Table 1, both measurements are in excellent
agreement (see $\S$3.2 for a comparison between \ion{Fe}{2} lines),
and they also agree very well with the predicted 
$W_\lambda$ of the \ion{O}{1} triplet given by Asplund
et al. (2004, hereafter Asp04).

Employing a Kurucz overshooting model atmosphere (Castelli, Gratton \& Kurucz 1997) 
and $W_\lambda$ from both the K84 (Hinkle et al. 2000) and VLT/UVES solar spectrum we found
A(O) = 8.65 ($\sigma$ = 0.03).
This is in very good agreement to other recent NLTE determinations of the
oxygen abundance using the \ion{O}{1} triplet. For example,
Mel\'endez (2004) found A(O) = 8.68 employing a spatially and temporally 
averaged 3D solar model, and A(O) = 8.67 with a Kurucz solar model.
Full 3D + NLTE calculations by Asp04 and Allende Prieto et al. (2004) 
resulted in A(O) = 8.64 and 8.70, respectively,
while A(O) = 8.70 was obtained by Shchukina et al. (2005) 
with 1.5D NLTE calculations in a 3D solar model (Asplund et al. 2000).
The recommended solar oxygen abundance by Asp04 is A(O) = 8.66 $\pm$ 0.05,
which is based on 3D analyses of the forbidden, permitted and infrared
OH lines of the $\Delta v$ = 0, 1 sequences. If we also consider the solar
oxygen abundance obtained from the first-overtone OH lines, then
the weighted mean solar O abundance is 8.64 (Mel\'endez 2004). 
The value adopted in this work is A(O)$_\odot$ = 8.65.

\subsection{Fe II lines}
As previously discussed in the literature 
(e.g. Lambert et al. 1996; Asplund et al. 2000; Gehren, Korn \& Shi 2001),
there is a lack of accurate experimental $gf$-values for \ion{Fe}{2} lines.
With the advent of 8-10 m telescopes and very efficient
spectrographs, high S/N high spectral resolution data
can be readily obtained, and therefore one of the
main limitations for accurate stellar abundance work
is the uncertainty in the transition probabilities of \ion{Fe}{2} lines.

In the present work the $gf$-values of \ion{Fe}{2} lines are from a revised version
of the list of Mel\'endez \& Barbuy (2002), where relative $gf$-values
within a given multiplet were taken from theoretical determinations
and the absolute scale of the transition probabilities of each multiplet
was determined from laboratory lifetimes and branching ratios. 
The approach of Mel\'endez \& Barbuy (2002) has the advantage 
of improving the accuracy of the $gf$-values
while preserving the laboratory scale of the oscillator strengths. 
We stress here that
a single correction factor can not be applied to the
whole set of theoretical $gf$-values, since the theoretical 
calculations should only be reliable within 
a single multiplet. This is why each multiplet needs to be
corrected individually, in order to put the whole set of 
multiplets into the absolute laboratory scale.

The revised \ion{Fe}{2} line list takes into account new
laboratory measurements by Schnabel et al. (2004) and
new theoretical calculations by R. L. Kurucz\footnote{http://kurucz.harvard.edu/}.
Further details and the complete list of \ion{Fe}{2} lines 
for the optical region will be presented in Mel\'endez \& Barbuy (2006). 
The subset of $gf$-values used for the
solar and stellar analysis are given in Table 2.
The interaction broadening constants ($C_6$) of the
\ion{Fe}{2} lines were obtained from the cross sections 
recently computed by Barklem \& Aspelund-Johansson (2005).

As an example of the quality of the new $gf$-values, 
in Fig. 1 are shown the solar iron abundances obtained from \ion{Fe}{2} lines 
by Hannaford, Lowe \& Grevesse (1992, hereafter H92), and the
rescaled abundances 
(A(Fe)$^{\rm new}$ = A(Fe)$^{\rm H92}$ + log {\it gf}$^{\rm H92}$ - log {\it gf}$^{\rm new}$) 
due to the new $gf$-values given in the present work.
As can be seen, significant progress
has been achieved, reducing the scatter 
from $\sigma$ = 0.07 dex ($gf$-values from H92) to $\sigma$ = 0.03 dex 
(new $gf$-values). The mean solar iron abundance from
H92 is 7.47, and with the improved $gf$-values is 7.46.
Similar comparisons with other works in the
literature also show that our new $gf$-values lead
to a decrease in the line-to-line scatter of the solar iron abundance 
(Mel\'endez \& Barbuy 2006).

It is also interesting to compare our new $gf$-values of
\ion{Fe}{2} lines with the solar $gf$-values determined
by Gurtovenko \& Kostik (1989, hereafter GK89), which should be reliable in a relative scale,
hence allowing to check our relative (theoretical) oscillator strengths.
A comparison between 20 lines in common shows an
excellent relative agreement, with a $\sigma$ = 0.038 dex, but
with a systematic difference of 0.18 dex between both sets,
that is mainly due to the high solar Fe abundance
adopted by GK89.

In order to determine the solar iron abundance
using our new $gf$-values of \ion{Fe}{2} lines, 
we selected \ion{Fe}{2} lines with the cleanest profiles in
the NSO FTS solar flux spectrum (Hinkle et al. 2000; K84).
As previously noted by H92, the lines at 5525.1 and 5627.5 \AA\
are significantly perturbed at the wings. H92 included these lines
by fitting Gaussian lineshapes, however, we discarded them because
the deblended profiles were significantly asymmetric.
Our measured $W_\lambda$ from the NSO solar flux spectrum, 
which are reported in Table 2,
are in excellent agreement with our measurements from 
the VLT/UVES reflected solar spectrum ($\S$3.1),
with a mean difference (K84 - UVES) of only $-$0.1 $\pm$ 1.2 m\AA\ 
(= $-$0.1 $\pm$ 2.8\%).

Our \ion{Fe}{1} + \ion{Fe}{2} atomic model for the
NLTE calculations includes over 250 levels
and nearly 500 radiative transitions. The model is similar to that
used by Shchukina \& Trujillo Bueno (2001) and Shchukina et al. (2003, 2005).

Employing the Kurucz model atmosphere we obtained a solar
iron abundance A(\ion{Fe}{2})$_{\rm NLTE}$ = A(\ion{Fe}{2})$_{\rm LTE}$ = 7.45 ($\sigma$ = 0.04). 
This value is in good agreement with 1.5D+NLTE calculations in a 3D 
solar model (Asplund et al. 2000) by Shchukina \& Trujillo Bueno (2001),
who found A(\ion{Fe}{1}) = 7.50, and also in good agreement with
previous 1D and 3D results from \ion{Fe}{2} lines by H92,
Schnabel, Kock \& Holweger (1999), and Asplund et al. (2000),
who obtained A(\ion{Fe}{2}) = 7.47 (1D), 7.42 (1D) and 7.45 (3D), respectively. 
We adopted A(\ion{Fe}{2})$_\odot$ = 7.45.

We end this section by showing the effect of different model
atmospheres on the solar iron abundance and its $\sigma$, employing 
3 different sets of $gf$-values for 13 \ion{Fe}{2} lines 
in common between the present work, H92 and GK89.
The calculations were performed using the same set
of input equivalent widths (Table 2) and v$_t$ = 0.9 km s$^{-1}$. 
We employed the Kurucz overshooting model,
a MARCS model (Asplund et al. 1997), a
spatially and temporally averaged 3D solar model atmosphere ($<$3D$>$, Asp04),
and the Holweger \& M\"uller (1974, hereafter HM74) solar model.
The results are shown in Table 3. As can be seen,
the Fe abundance obtained from \ion{Fe}{2} lines has
only a small dependence on the adopted model atmosphere. 
A line-to-line abundance scatter of $\sigma \approx$ 0.05 dex
is obtained for our $gf$-scale and the solar $gf$-scale of GK89,
while the worst scatter (0.11 dex) is obtained with
the H92 $gf$-values. Certainly, $\sigma$ can not be used
as the only criterion for the quality of a $gf$-scale,
since the scatter depends on both the adopted model atmosphere
(see Table 3) and v$_t$ (see Gehren et al. (2001) and 
Kostik, Shchukina \& Rutten (1996) for details on the effects of v$_t$), 
but a good $gf$-scale should give a small $\sigma$
in order to give consistent results when only few lines are available for analysis.

\section{Reddening}

Before colors are used to determine \tsin, they have to be corrected 
for interstellar absorption. A good determination of the reddening
is specially important for metal-poor turn-off stars
due to the steep slopes of their temperature vs. color relations.
$E_{B-V}$ values were estimated employing:
{\it i)} several reddening maps ($E_{B-V}^{\rm maps}$);
{\it ii)} relative stellar flux distributions ($E_{B-V}^{\rm SED}$);
and {\it iii)} Str\"omgren photometry ($E_{B-V}^{\rm uvby-\beta}$),
as given by N04. Each method is described below.

\subsection{Reddening maps}

Several studies show that nearby stars closer than 75 pc
have negligible reddening, since most stars within this distance are located
inside the ``Local Bubble'' of radius $\approx$ 70-75 pc 
(e.g. Lallement et al. 2003; Breitschwerdt et al. 2000;
Sfeir et al. 1999; Leroy 1999; Vergely et al. 1998).
\footnote{Note, however, that the Local Bubble is by no means spherical,
but roughly within 75 pc of the Sun the extinction is negligible,
although in some cases even stars as far as 100 pc can be almost
unreddened}

We consider several extinction surveys 
(e.g. Fitzgerald 1968; Neckel \& Klare 1980; Arenou et al. 1992, hereafter A92) 
included in a Fortran code by Hakkila et al. (1997, hereafter H97), 
adopting the weighted-average with the inverse square of the errors as weights. 
The A92 extinction model included in the H97 code
seems to systematically overestimate the reddening for stars with a distance
$d$ $<$ 0.5 kpc, as shown by Chen et al. (1998, hereafter C98), who found
that the average extinction derived from Str\"omgren photometry
is about 40\% lower than the average extinction derived from the A92 maps. 
Instead of discarding $E_{B-V}^{A92}$ values for stars closer than 500 pc,
we doubled its error (equivalent to lower its weight by a factor of 4)
and corrected $E_{B-V}^{A92}$ by a factor of 0.6 for stars in this distance range,
as suggested by the results of C98.

We also employ the empirical reddening laws by Bond (1980, hereafter B80) 
and C98, which are both cosecant laws dependent on distance and Galactic latitude, 
although for low latitude ($|$b$|$ $\leq$ 10$^{\circ}$) objects closer than 1 kpc
C98 law also includes a dependence with Galactic longitude. 

Reddening corrections were also obtained 
from the $E_{B-V}$ map by Schlegel et al. (1998,
hereafter S98), which is based on COBE/DIRBE and IRAS observations. 
Note that this map seems to systematically overestimate the 
reddening by about 20\% 
(e.g. Arce \& Goodman 1999; Chen et al. 1999; Dutra et al. 2003a,b),
and some studies argue for a systematic zero point error in S98 map. 
For example, Hudson (1999) found that the
$E_{B-V}$ obtained from S98 is about 0.016 mag higher than 
the reddening of 86 RR Lyrae given by Burstein \& Heiles (1978).
We have made the same comparison for the sample of RR Lyrae
given in Hudson (1999), and have found that indeed
for $E_{B-V}$ $<$ 0.2, $E_{B-V}^{S98}$ should be corrected by $-$0.02 mag.
Burstein (2003) confirmed the existence of a 
zero point difference of $-$0.02 mag between the maps of S98 and
those from Burstein \& Heiles (1978).

We adopted the following correction for the S98 maps:

$E_{B-V}^{S98c}$ = 0.9 $E_{B-V}^{S98}$ - 0.01.

The different prescriptions for the determination of
$E_{B-V}^{\rm maps}$ were combined as follows:

\begin{itemize}
\item $E_{B-V}^{\rm maps}$ = 0.0 for stars within 75 pc.

\item For stars with 75 pc $< d \leq$  100 pc, 
$E_{B-V}^{\rm maps}$ = 
($E_{B-V}^{H97}$ + ($E_{B-V}^{B80}$ + $E_{B-V}^{C98}$)/2 + 2 $\times$ 0.0)/4.
The factor 2 $\times$ 0.0 considers a 50 \% chance of 
zero reddening.

\item If a star has $d >$ 100 pc and $|$b$|$ $>$ 45 $^{\circ}$,
$E_{B-V}^{\rm maps}$ = 
( $n$ $E_{B-V}^{H97}$ + $m$ $E_{B-V}^{S98c}$)/($n+m$), 
where $n$ is the square root of the number of maps used in 
H97 and $m$ = ($d$+1.5)(sin$|$b$|$)$^{1.5}$, with $d$ in kpc. 
This empirical parametrization gives a high weight to the S98 map for 
high-latitude and/or distant objects, and a weight essentially
zero for low-latitude stars.

\item Stars with $d >$ 100 pc and $|$b$|$ $\leq$ 45 $^{\circ}$,
$E_{B-V}^{\rm maps}$ = 
( $n$ $E_{B-V}^{H97}$ + $m$ $E_{B-V}^{S98c}$ + 
($E_{B-V}^{B80}$ + $E_{B-V}^{C98}$)/2)/($n+m+1$). 
The weights $n$ and $m$ are equal to the previous case,
except that $m$ = 0 if $E_{B-V}^{S98c}$ $>$ 2$E_{B-V}^{H97}$ 
(in some cases this restriction was relaxed, especially for 
very distant objects with $|b| > 30^{\circ}$).

\item If after applying the above criteria 
$E_{B-V}^{\rm maps}$ $>$  $E_{B-V}^{S98c}$, then
we adopted $E_{B-V}^{\rm maps}$ = $E_{B-V}^{S98c}$.

\end{itemize}

The distances employed here were estimated from Hipparcos parallaxes, 
Str\"omgren photometry (N04) and isochrones (see $\S6$).

\subsection{Extinction from Stellar Energy Distributions}

The observed relative spectral energy distribution (SED) of a star 
is defined primarily by its \teff and by interstellar extinction. The 
right $E_{B-V}$ value can be used to correct the observed SED, 
recovering thus the unredenned SED. Hence, 
given several observed colors (X-Y)$_i$ and highly accurate 
color vs. \teff relations, the minimum scatter between 
different color temperatures (\tsin(X-Y)$_i$)
should be obtained for the right choice of $E_{B-V}$,
denoted by $E_{B-V}^{\rm SED}$.

We employed our IRFM \teff scale (RM05b),
which includes seventeen colors from the blue to the near
infrared, allowing for a good estimate of $E_{B-V}^{\rm SED}$.
Considering that our \teff calibrations have high internal accuracy,
the main uncertainty in the determination of $E_{B-V}^{\rm SED}$
arises from photometric errors. We typically employed eleven colors
for each sample star ($\S5$), alleviating in this way the impact of
photometric errors.

\subsection{Extinction from Str\"omgren Photometry}

N04 determined interstellar reddening 
for the sample stars employing Str\"omgren $uvby$-$\beta$ photometry
and the calibration by Schuster \& Nissen (1989),
including a zero-point correction of +0.005 mag (Nissen 1994).

The $E_{b-y}$ given by N04 was transformed to $E_{B-V}$
applying: $E_{B-V}^{\rm uvby-\beta}$ = 1.35 $E_{b-y}$ (Crawford 1975).
When the $E_{b-y}$ value given by N04 was negative
we adopted $E_{B-V}^{\rm uvby-\beta}$ = 0.0.

\subsection{Comparisons and Adopted Reddening}

In Fig. 2 are compared the different $E_{B-V}$ values
determined above. 
There are not clear correlations between the different methods.
Fortunately, most of the points (0.005 $<$ $E_{B-V}$ $<$ 0.030)
are randomly distributed along the 1:1 line, which means
that a correlation may exists, but it is hidden by
the errors in the reddening determinations.
A larger sample, including very reddened stars (at least
up to $E_{B-V}$ = 0.1 mag) is necessary in order to
evaluate the different methods employed to estimate
the reddening.

The mean differences between the methods are: 
$E_{B-V}^{\rm maps}$ - $E_{B-V}^{\rm SED}$ = 0.002 mag ($\sigma$ = 0.011 mag);
$E_{B-V}^{\rm maps}$ - $E_{B-V}^{\rm uvby-\beta}$ = 0.004 mag ($\sigma$ = 0.012 mag);
$E_{B-V}^{\rm uvby-\beta}$ - $E_{B-V}^{\rm SED}$ = 0.000 mag ($\sigma$ = 0.015 mag).
Considering that there are not significant zero point differences
between the different methods, we adopted the mean value:

$$E_{B-V}^{\rm adopted} = (E_{B-V}^{\rm maps} + E_{B-V}^{\rm SED} + E_{B-V}^{\rm uvby-\beta})/3$$ 

In Table 4 are shown the adopted $E_{B-V}$ values
and the scatter $\sigma$ between the three methods 
(typically $\sigma$ = 0.008 mag).

\section{Effective Temperatures}

We used the new IRFM temperature scale of RM05b, 
which is based on a homogeneous
analysis of more than 10$^3$ stars, 
for which IRFM temperatures were obtained
employing updated atmospheric parameters
(RM05a). The main improvements compared with previous works 
are a better coverage of the atmospheric parameters
space (\tsin, log $g$, [Fe/H]), the use of up-to-date 
metallicities, and the fit of trends in the residuals,
thus eliminating any spurious trend in 
the \tsin:color:[Fe/H] relations (RM05b).
The use of updated metallicities and the good
coverage of the atmospheric parameters space
were crucial to derive reliable \teff calibrations, 
greatly helping to distinguish noise from real trends with 
metallicity.
Seventeen colors were calibrated in the UBV, {\it uvby}, Vilnius,
Geneva, RI (Cousins), DDO, Tycho (Hipparcos), and 2MASS 
photometric systems (RM05b). 

The colors of the sample stars were mainly
obtained from the General Catalogue of Photometric Data
(Mermilliod et al. 1997), the Hipparcos/Tycho Catalogue (ESA 1997),
and the final release of the 2MASS Survey (Cutri et al. 2003).
Almost all stars in the sample have $B-V$, $b-y$, Geneva and 2MASS colors,
and, when available, we also used Vilnius, Cousins, DDO and Tycho colors.
At least four colors were used for the temperature determination,
although for most stars eleven colors were available. 
The mean, weighted average (using the error of each color
calibration as weight) and trimean
\footnote{the trimean $T$ is a robust estimate of central tendency.
We adopted Tukey's trimean $T$ = (Q1 + 2 $\times$median + Q3)/4,
where Q1 and Q3 are the first and third quartile} 
temperatures were computed, as well as the standard deviation, 
weighted error and quartile deviation. 
In general all estimates agree very well, except when
an outlier was present. 
We adopted the trimean and
the pseudo standard (quartile) deviation, which is the robust equivalent of $\sigma$ 
in a normal distribution. \footnote{the pseudo standard deviation $\sigma$
was obtained from the quartile deviation QD (=(Q3-Q1)/2), employing
$\sigma$ = 3/2 QD}
The adopted temperatures and $\sigma$ are shown in Table 4.

\section{Gravities}

Good ($\pi/\sigma(\pi) >$ 4.5) Hipparcos parallaxes  were
employed to obtain trigonometric surface gravities (log $g$(Hip)) 
with errors (only due to $\pi$) lower than 0.2 dex in log $g$. The stellar mass
was obtained from enhanced $\alpha$-element 
Y$^2$ isochrones (Kim et al. 2002; Demarque et al. 2004), by finding
the closest isochrone to the \tsin(IRFM)/Luminosity(Hipparcos)/[Fe/H](\ion{Fe}{2})
of the sample star. In this way we simultaneously determined 
isochronal ages and masses. Typical ages and masses are about 11 Gyr and
0.82 M$_\odot$, respectively. The use of other isochrones (e.g. Padova
or Victoria) result in similar masses ($\approx$ 0.8 M$_\odot$ for a
12 Gyr turn-off at [Fe/H] = $-$2.3), but we have adopted the
Y$^2$ isochrones because they extend to [Fe/H] = $-$3.3,
while most Padova and Victoria isochrones are available
only for [Fe/H] $\gtrsim$ $-$2.3.

We also estimated surface gravities from Y$^2$ isochrones
(log $g$(Y$^2$)). For turn-off stars only one solution is
found at a given \tsin, but for other stars the solution
is twofold. When good parallaxes were
available the solution closer to log $g$(Hip) was chosen,
otherwise we used as additional constraints {\it i)} photometric $M_V$ as 
determined from Str\"omgren photometry by N04, 
{\it ii)} Hipparcos parallaxes (when $\pi/\sigma(\pi) >$ 2), and 
{\it iii)} previous log $g$ data given in the literature.
In some cases even when all these constraints were
used it was still unclear (within the errors) whether the star 
was below or above the turn-off.
In those cases we adopted as a compromise 
log $g$(Y$^2$) = log $g$(turn-off), 
and due to this uncertainty 
in evolutionary status the error in log $g$ (Y$^2$) for those stars 
is about 0.25 dex (this is the maximum possible log $g$ shift
from the turn-off to both the MS and the subgiant branch).

For about half of the sample we have available good
Hipparcos parallaxes, hence we can check our
isochronal gravities. As shown in Fig. 3,
both agree very well, with a mean difference 
log $g$(Y$^2$) - log $g$(Hip) = $-$0.015 ($\sigma$ = 0.090).
Based on this comparison, we adopted an error of 0.15 dex for log $g$(Y$^2$),
except when there was a large uncertainty in
evolutionary status, in which case an error
of 0.3 dex was assigned.

The adopted surface gravity is the weighted average 
of log $g$(Y$^2$) and log $g$(Hip),
and the adopted error is $\sigma_{{\rm log} g}$ = max(0.1 dex, $\sigma$, $\sigma_{weighted}$).
These surface gravities and 1-$\sigma$ errors are given in Table 4.

\section{Abundances}

We employed Kurucz overshooting model 
atmospheres\footnote{http://kurucz.harvard.edu/}
with 72 layers (Castelli et al. 1997), adopting a metallicity [M/H] (Table 4) 
about 0.2 dex higher than the iron abundance [Fe/H] obtained
in the literature (including the results obtained in this work), 
in order to compensate for the enhancement of 
$\alpha$-elements in halo stars (e.g. Sneden et al. 1994; Fulbright \& Kraft 1999). 
The use of solar-scaled Kurucz models instead of
alpha-enhanced models have negligible impact on the
[O/Fe] ratio derived in the present work (see $\S$7.1 and
Table 5, where we show the sensitivity of [O/Fe] to [M/H]).

The calculations were performed in LTE and NLTE employing the 
code NATAJA, which is described in Shchukina \& Trujillo Bueno (2001) and
Shchukina et al. (2005). The adopted atomic data for oxygen and
iron were described in $\S$3.
LTE computations were also done with the latest version of MOOG 
(Sneden 1973). Both LTE computations agree very well, typically within 
0.015 and 0.025 dex for \ion{Fe}{2} and \ion{O}{1}, respectively.
The small differences are probably due to different continuum opacities
adopted in the codes.

For stars with [Fe/H] $\geq$ $-$2 we determined
microturbulence velocities v$_t$ by requiring no dependence 
of [Fe/H] against reduced equivalent width; we found a typical
$v_t$ = 1.5 km s$^{-1}$.
For stars with [Fe/H] $<$ -2 the \ion{Fe}{2} lines and \ion{O}{1} triplet 
are very weak and essentially independent of microturbulence (Table 5). 
For these very metal-poor stars we adopted $v_t$ = 1.5 km s$^{-1}$ (N04).

In Table 4 are given the oxygen and iron abundances derived 
in the present work, as well as the line-to-line scatter (i.e., errors mainly
due to errors in $W_\lambda$ and $gf$-values).
The [\ion{O}{1}/\ion{Fe}{2}]$_{\rm NLTE}$ ratios obtained in this work
are plotted in Fig. 4a,b,c, as a function of [Fe/H], \teff and log $g$,
respectively. 

In this work we obtained a mean [O/Fe] = 0.49 dex (average)
and a weighted mean [O/Fe] = 0.50 dex (see $\S$7.2).

\subsection{Overshooting vs. no overshooting}

We have assessed the effect of using the latest no
convective overshooting (NOVER) Kurucz models 
(Castelli \& Kurucz 2003) instead of the 
overshooting models (Castelli et al. 1997) adopted in the
present work.

The new NOVER Kurucz models adopt new ODFs, 
and models with both solar-scaled and $\alpha$-enhanced abundances are available
in the range -2.5 $\leq$ [Fe/H] $\leq$ +0.5. 

The solar-scaled NOVER models
with a metallicity [M/H] = [Fe/H] + 0.2 dex resulted
in essentially the same (within 0.01 dex) oxygen and iron 
abundances than those obtained with the $\alpha$-enhanced NOVER models,
which supports our choice of [M/H] for the model atmospheres.

The use of the new NOVER models (both solar-scaled and $\alpha$-enhanced)
gives lower abundances compared to overshooting models. 
The solar oxygen and iron abundances 
obtained from \ion{Fe}{2} and \ion{O}{1} lines
are both reduced by 0.06 dex with
the new NOVER models, while for the sample
stars the Fe and O abundances are about 0.06-0.07 and
0.08-0.10 dex lower, respectively.
Therefore, the use of the latest
NOVER Kurucz models does not affect the results presented here,
since the [Fe/H], [O/H] and [O/Fe] ratios derived with the 
NOVER models are roughly preserved. 
Indeed, there is a small reduction of about 0.02 dex in
[O/Fe] when the new Kurucz models without convective
overshooting are employed, reinforcing
thus the low [O/Fe] ratios found in the present work. 

\subsection{Intrinsic Scatter and Errors}

We found a moderate star-to-star scatter in [O/Fe]
of $\sigma_{obs}$ = 0.136 dex.
The two main sources of observed star-to-star scatter 
are reddening and statistical errors (line-to-line scatter). 
The error in reddening introduces error in the temperature 
(and to some extent also in log $g$), thus affecting the
oxygen (and iron) abundance.
An uncertainty of 0.01 mag in $E_{B-V}$ corresponds to an 
uncertainty of $\pm$ 52 K in \tsin. 

In Fig. 4 it is shown that the stars with 
an uncertainty $\Delta$$E_{B-V}$ $>$
0.01 mag (plotted as open circles) are the ones that deviate the most
from the mean [O/Fe].
In fact, while 54\% of the sample stars with $\Delta$$E_{B-V}$ $>$ 0.01 mag
(open circles, Fig. 4) deviate by more than 0.1 dex from 
[O/Fe] = +0.5 dex, only 28 \% of the sample stars with lower reddening 
uncertainties (filled circles, Fig. 4) show the
same discrepant behavior.
Considering only the stars with $\Delta$$E_{B-V}$ $\leq$
0.01 mag, the observed scatter in [O/Fe] is reduced to $\sigma$ = 0.10 dex.


Other sources of errors are due to uncertainties in the
parameters of the model atmosphere (\tsin, log $g$, v$_t$, [M/H]).
In Table 5 are shown the impact of errors 
for changes of $\Delta$\teff = 50 K, $\Delta$log $g$ = 0.15 dex,
$\Delta$v$_{\rm t}$ = 0.3 km s$^{-1}$, and $\Delta$[M/H] = 0.2 dex.

For each star in the sample we estimated the error in [O/Fe]
considering: {\it i)} the uncertainty in reddening given in column 2 of Table 4,
adopting an error of 0.04 dex per 0.01 mag, due to its impact on \tsin;
{\it ii)} the uncertainties in \teff and log $g$ given in columns 3 and 4 of Table 4;
{\it iii)} an error of 0.3 km s$^{-1}$ in microturbulence;
{\it iv)} and uncertainty of 0.2 dex in [M/H] of the model
atmosphere; and 
{\it v)} statistical uncertainties 
(line-to-line scatter) in the abundances of iron and oxygen
given in columns 9 and 12 of Table 4, respectively. The 
error in [O/Fe] is shown in the last column of Table 4.
The weighted mean (adopting the inverse square of the
errors as weights) of the oxygen-to-iron ratio is
[O/Fe] = +0.50 dex.

The mean error in [O/Fe] is
0.121 dex (trimean = 0.112 dex), which is 0.015 dex
lower than the observed star-to-star scatter in [O/Fe]
($\sigma_{obs}$  = 0.136 dex). 
This means that the errors were slightly underestimated, or that
there is some small intrinsic scatter in the sample.

In addition to the errors reported in Table 4, 
there may be systematic errors in our analysis due to errors in the IRFM
\teff scale, NLTE effects, and granulation effects (e.g. Asplund 2005). 

\section{Discussion}

Since the equivalent widths used in this work were taken
from Ake04 and N04, we first compare our results to theirs ($\S$8.1),
then we discuss previous studies, 
especially those works claiming a step 
increase in [O/Fe] for lower metallicities based on 
the \ion{O}{1} triplet ($\S$8.2) and UV OH lines ($\S$8.3).
We also compare our results with previous works that obtained a flat [O/Fe] 
based on hot \teff scales ($\S$8.4). Then, we discuss whether
high effective temperatures for metal-poor turn-off stars 
are physically reasonable, as well as the \teff of
hyper metal-poor turn-off stars ($\S$8.5).

\subsection{Comparison with Ake04/N04}

The main differences between our work and that of Ake04/N04
are: different \teff scale (and reddening), new set of $gf$-values for \ion{Fe}{2},
NLTE calculations for both \ion{Fe}{2} and \ion{O}{1}, and the
use of different model atmospheres. Note that the NLTE corrections
$\delta_{\rm NLTE} (\equiv$ A$^{\rm NLTE}$ - A$^{\rm LTE}$)
for \ion{O}{1} used by Ake04 were interpolated from previous
calculations by Nissen et al. (2002), while in our case we
explicitly computed the LTE and NLTE abundances for each star.

In Fig. 5 are shown the differences between the present work
and that of Ake04/N04, as a function of 
[Fe/H] (left panels) and \teff (right panels).
In Figs. 5b,c and 5f,g an outlier can be seen at 
[Fe/H] $\approx -$0.7 and \teff $\approx$ 5650 K, respectively.
This is probably due to a typo in the oxygen abundance of 
BD +08 3065 (G 016-013) given by Ake04, since its O abundance 
can not be lower than that of HD 146296, which has both 
lower $W_\lambda$ and higher \teff than BD +08 3065.

The differences $\Delta$[Fe/H] and $\Delta$A$_{Fe}^{\rm LTE}$ are shown in Figs. 5a,e
with filled and open circles, respectively. Despite our higher
\teff (as shown in Figs. 5d,h), our A$_{Fe}^{\rm LTE}$ abundances 
are in excellent agreement with Ake04/N04. This can be explained by the low 
sensitivity of the iron abundance to changes in \teff (Table 5).
The difference in [Fe/H] is independent of [Fe/H] and \tsin, but
with an offset of 0.11 dex, which is due to the use of 
different model atmospheres, solar abundances\footnote{The
solar Fe abundance was not determined by
Ake04/N04, but it should be in the same system
of Nissen et al. (2002), where A(Fe)$_\odot$ = 7.53
was found.}, $gf$-values and NLTE corrections
(Ake04/N04 do not correct for NLTE effects,
while our $\delta_{\rm NLTE}$(\ion{Fe}{2}) $\approx$ 0.046 dex). 

Our [O/Fe]$_{\rm NLTE}$ ratios are lower than those in Ake04, as can be
seen in Figs. 5b,f. The difference $\Delta$[O/Fe]$_{\rm NLTE}$ depends
on metallicity, reaching about $-$0.3 dex at [Fe/H] = $-$3.
This is partly explained by our higher \teff for lower metallicities (Figs. 5d,h),
but also for the small NLTE corrections used by Ake04 for
the most metal-poor stars. 
In Figs. 5c,g the differences in LTE oxygen abundance ($\Delta$A$_{O}^{\rm LTE}$)
 and in NLTE corrections ($\Delta \delta_{\rm NLTE}$)
are represented by open circles and stars, respectively.
There is a good agreement in the A$_{O}^{\rm LTE}$,
although for the most metal-poor stars we expected to see
a larger discrepancy in A$_{O}^{\rm LTE}$ due to the lower \teff adopted by Ake04. 
The NLTE corrections are similar for [Fe/H] $> -$2.5, with an small
offset of $-$0.053 dex. For lower metallicities our NLTE corrections
are considerably larger, 
being about 0.25 larger (more negative) at [Fe/H] $\approx -$3.

The interpolated $\delta_{\rm NLTE}$ for the five
stars with the lowest metallicities have been
wrongly estimated by Ake04 (M. Asplund, 2005, private communication). In fact, 
the original NLTE correction by Nissen et al. (2002) for LP 815-43, 
the star with the lowest metallicity ([Fe/H] = $-$2.7) in their
sample, is $\delta_{\rm NLTE}$(\ion{O}{1}) = $-$0.25 dex, 
which is in excellent agreement with our $\delta_{\rm NLTE}$(\ion{O}{1}) = $-$0.24 dex.

\subsection{Linear Relationship vs. Flat [O/Fe]}

As can be seen in Figs. 4a,b,c,
the [O/Fe] ratio is flat ([O/Fe] $\approx$ +0.5) and independent
of metallicity, temperature and surface gravity, respectively.
On the other hand, some previous investigations of the
\ion{O}{1} triplet have found a steep linear relationship
between [O/Fe] and [Fe/H] (Abia \& Rebolo 1989;
Cavallo et al. 1997; Israelian et al. 1998, 2001; Mishenina et al. 2000),
all with a similar slope of about $-$0.35,
and reaching [O/Fe] $\approx$ 1.1 at [Fe/H] = $-3$.
Interestingly, the work by Tomkin et al. (1992)
found a flat [O/Fe], but with a high mean [O/Fe] = 0.8.
Note also that the analysis of the \ion{O}{1} triplet by
Nissen et al. (2002) shows only a mild dependence of [O/Fe] with 
metallicity, with [O/Fe] $\approx$ 0.4 at [Fe/H] = $-1$, and
increasing to [O/Fe] $\approx$ 0.6 at [Fe/H] = $-2.5$. The
[O/Fe] ratios obtained by Ake04 are similar to those
obtained by Nissen et al. (2002), showing a mild dependence with
metallicity and reaching [O/Fe] $\approx$ 0.7 at [Fe/H] = $-3$.
However, as we have seen in $\S$8.1, the mild slope found 
by Ake04 can be reduced by employing our \teff scale and also
proper NLTE corrections for [Fe/H] $< -$2.5.

For comparison purposes between our results and previous studies
which have found a steep linear trend between [O/Fe] and [Fe/H],
a typical linear relationship found by those works
is shown with a dotted line in Fig. 4a. 
Clearly, our results do not  support previous claims for a large
increase of [O/Fe] for decreasing metallicities.

Besides the neglect (or underestimation) of NLTE effects for \ion{O}{1}, 
the main reason for the large slope found in other works
based on the \ion{O}{1} triplet seems to be the lower \teff adopted. 
For example, we show in Fig. 6 the difference $\Delta$\teff between the IRFM temperatures
obtained with the \teff scale by RM05b 
and those adopted by Israelian et al. (1998, 2001). 
The difference $\Delta$\teff increases with 
decreasing [Fe/H], reaching about $\Delta$\teff = +350 K at
[Fe/H] = $-3$, thus leading Israelian et al. (1998, 2001)
to derive much higher (and metallicity-dependent) [O/Fe] ratios.

A better understanding of the differences between our
temperatures and those adopted by Israelian et al. (1998, 2001)
can be achieved by examining the differences between the
temperature scales adopted by them and us (RM05b).
Israelian et al. (1998) adopted the {\it b-y} and {\it V-K} 
\teff calibration by Alonso et al. (1996), 
while Israelian et al. (2001) adopted 
{\it V-K} calibrations by Alonso et al. (1996) and Carney (1983).
In Fig. 7 we show the differences between the {\it b-y} and {\it V-K}
calibrations by RM05b and those by Alonso et al. (1996) and Carney (1983).

In their first paper, which mainly 
analyzes stars cooler than 6000 K, Israelian et al. (1998)
adopted the mean of the {\it b-y} and {\it V-K} calibrations
by Alonso et al. (1996). As can be seen in Fig. 7, for \teff $<$
6000 K, the {\it b-y} and {\it V-K} Alonso et al. (1996)
temperatures are higher and lower than RM05b, respectively.
Thus the mean color temperatures are similar to RM05b, 
explaining the good agreement for \teff $<$ 6000 K
(Fig. 6). 

In their second paper, which deals
mainly with very metal-poor ($-3.4 <$[Fe/H]$< -2.5$) 
stars hotter than 6000 K, Israelian et al. (2001)
adopted the mean from the {\it V-K} calibrations by
Alonso et al. (1996) and Carney (1983). Fig. 7 shows that
for \teff $>$ 6000 K the {\it V-K} calibration by Alonso et al.
is lower than that of RM05b by about 180 K, while the 
{\it V-K} calibration by Carney et al. (1983) is lower than
RM05b by about 340 K in the relevant \teff and [Fe/H] range
used by Israelian et al. (2001). So, on average, the calibrations
used by Israelian et al. (2001) are about 250 K 
lower than those by RM05b. 
Besides, Israelian et al. (2001) neglected reddening corrections, 
which are important for distant early G/F dwarfs. For stars hotter than
6000 K, we obtained a mean $E_{\it B-V}$ = 0.016 mag, which is
equivalent to $\Delta$\teff $\approx$ 85 K. So, considering altogether the
neglect of reddening (85 K) and the use of the Alonso et al. (1996)
and Carney (1983) calibrations (250 K), Israelian et al. (2001)
temperatures are lower than RM05b by about 335 K,
which explain the large difference shown in Fig. 6 for
\teff $>$ 6000 K.

A similar reasoning could be applied to explain the
differences between the temperatures obtained from
the calibrations by RM05b and those determined
by other authors. We also show in Fig. 7
the difference between our \teff scale and that of King (1993),
which was used by Boesgaard et al. (1999), who also used the
\teff scale by Carney (1983). 
It is important to mention
that not all works which have found large [O/Fe] ratios
are based entirely on photometric temperatures.
For example, part of the sample analyzed by Mishenina et al. (2000)
have temperatures determined from the fitting of H$\alpha$ line
profiles, and the other part was taken 
from the work by Cavallo et al. (1997), which is
a mix of temperatures obtained from the excitation
equilibrium of \ion{Fe}{1} lines, photometric temperatures
from the King (1993) \teff scale, temperatures from hydrogen profiles
by Axer, Fuhrmann, \& Gehren (1994), or average \teff from several values
reported in the literature. 

In principle, the temperature from
H$\alpha$ should be reliable, but in practice the H$\alpha$
temperatures are subject to problems in the continuum determination
(wings falling in different orders), as well as to the
treatment of convection (Castelli et al. 1997).
For example, Castelli et al. (1997) found a H$\alpha$
temperature of 6500 and 6700 K in Procyon, 
depending on the adopted prescription 
for convection. They obtained 
higher temperatures from other Balmer lines.
Interestingly, Aufdenberg, Ludwig \& Kervella (2005)
have recently found that Kurucz convective overshooting model
atmospheres may better represent the mean
temperature structure of F stars similar to Procyon, 
since high precision interferometric optical-red-infrared data 
of this star are consistent with a temperature structure 
with significant convective overshooting.

Although the absolute H$\alpha$ temperatures may be in error,
the relative temperatures should be reliable.
In Fig. 8 we show that the relative temperatures 
of metal-poor ($-3 <$ [Fe/H] $< -1$) turn-off stars 
determined by Asplund et al. (2005) 
using an accurate modeling of the H$\alpha$ line 
are in excellent agreement with those 
given by the IRFM \teff scale of RM05b, except
for the three more metal-poor stars ([Fe/H] $< -2.7$),
which have systematically lower H$\alpha$ temperatures.
Excluding those stars, 
the zero point difference (IRFM - H$\alpha$) is 
only 61$\pm$62 K, 
similar to the difference $\Delta$\teff = 34$\pm$95 K 
found by Asplund et al. (2005) between their H$\alpha$
temperatures and those from the IRFM \teff scale of Alonso et al. (1996).

\subsection{The new IRFM \teff scale and revised [O/Fe] from the UV OH lines}

Fields et al. (2005) have recently used effective temperatures from 
Mel\'endez \& Ram\'{\i}rez (2004), which are based on the 
new IRFM \teff scale by RM05a,b, 
in order to determine the oxygen abundance from the UV OH lines
in thirteen metal-poor dwarfs. 
Fields et al. (2005) find
very high [O/Fe] ratios, with a very 
steep slope: [O/Fe] = $-$0.66~[Fe/H] $-$ 0.25,
which is about twice as large as that claimed in
previous works based on the UV OH lines (Israelian
et al. 1998; Boesgaard et al. 1999). The slope
is reduced when the single point with [Fe/H] $< -3$ and an 
extreme [O/Fe] $\approx$ +2.4 is excluded, 
resulting in [O/Fe]$_{\rm 1D}^{\rm UV OH}$ = $-0.30~$[Fe/H] + 0.49,
which has a similar slope to that found
by Israelian et al. (1998) and Boesgaard et al. (1999),
but a higher constant term. Half of the higher constant
term is explained by the different solar abundances adopted by 
Fields et al. (2005; K. A. Olive \& B. D. Fields, 
private communication), leading to an increase of 
+0.23 dex in [O/Fe]. The other half is probably due
to the use of a hotter \teff scale.

As shown by Asplund \& Garc\'{\i}a P\'erez (2001),
oxygen abundances from the UV OH lines may be
severely overestimated by 1D analyses\footnote{Note that the large
3D abundance corrections apply mainly to hot turn-off stars, therefore
this is not necessarily in conflict with the recent low [O/Fe] ratios found in the 1D analysis
of UV OH lines in subgiants by Garc\'{\i}a P\'erez et al. (2005)}, and
the 3D abundance corrections depend on metallicity
{\it and} temperature. 
For example, for a star
with \teff $\approx$ 6000 K, the 3D correction 
$\Delta$[O/Fe]$_{\rm 3D - 1D}^{\rm UV OH}$ increases
from about $-0.5$ dex at [Fe/H] = $-2$, to $-0.7$ at [Fe/H] $\approx -3$.
On the other hand, at [Fe/H] = $-$3, increasing the
temperature from \teff $\approx$ 5890 to \teff $\approx$ 6200 K
increases the $\Delta$[O/Fe]$_{\rm 3D - 1D}^{\rm UV OH}$
correction from $-$0.5 to $-$0.9 dex (Asplund \& Garc\'{\i}a P\'erez 2001).
Therefore, the increase in the 1D oxygen abundance from the UV OH lines
due to higher temperatures is probably
compensated by the increase in the 3D corrections,
resulting in low [O/Fe] ratios.

The above reasoning is checked quantitatively in
Fig. 9, where we show the [O/Fe] ratios 
obtained by Fields et al. (2005) using the temperatures
by Mel\'endez \& Ram\'{\i}rez (2004), along with the
corresponding $\Delta$[O/Fe]$_{\rm 3D - 1D}^{\rm UV OH}$
corrections interpolated from 3D calculations 
by Asplund \& Garc\'{\i}a P\'erez (2001). 
Since 3D corrections are only available for
stars with \teff $\leq$ 6205 K, we refrained from
applying the corrections to the whole sample analyzed
by Fields et al. (2005), but only to stars
with \teff $<$ 6250 K (in fact, most stars shown
in Fig. 9 have \teff $<$ 6150 K).
The mean corrected [O/Fe] ratio is
[O/Fe]$_{\rm 3D}^{\rm UV \; OH}$ = 0.45.
Although it is true that our \teff scale
increases the [O/Fe] ratio obtained from the UV OH lines 
in a {\it 1D analysis}, it is also true that the 
3D corrections increase with higher \teff and lower [Fe/H], 
thus compensating the first effect and resulting in 
a roughly flat (and low) [O/Fe] ratio for halo stars (Fig. 9).

It is important to mention that we have assumed above
that the stellar parameters remain roughly unchanged 
with 3D model atmospheres. Shchukina et al. (2005) have
recently performed 1.5D+NLTE computations employing a
3D model atmosphere of the metal-poor subgiant HD 140283, 
and based on spectroscopic constrains
they suggested that the stellar parameters for this star
may need modification when 3D model atmospheres are employed. 
On the other hand, Asplund \& Garc\'{\i}a P\'erez (2001) show 
that when the effective temperatures of metal-poor stars 
of solar \teff are determined with the IRFM, only small 
changes in \teff (below 20 K) are expected in 3D. The
changes are even smaller (below 5 K) for metal-poor turn-off stars.
In their analysis of the UV OH lines, Fields et al. (2005)
adopted effective temperatures by Mel\'endez \& Ram\'{\i}rez (2004),
which are in the IRFM \teff scale of RM05b, and therefore 
those \teff should be little affected in 3D model atmospheres.
The main concern is probably the iron abundance, 
which for the only star (HD 140283) with a 1.5D+NLTE calculations
in a 3D model atmosphere (Shchukina et al. 2005), 
is 0.25 dex higher than in the present 1D analysis.
The higher metallicity would result in a lower 3D correction,
which is compensated by the increase of 0.25 dex in [Fe/H], 
resulting in slightly lower [O/Fe] ratios than those presented in Fig. 9. 
Note also that the UV OH lines may be affected by NLTE effects, 
perhaps increasing the oxygen abundances from UV OH lines
(Asplund \& Garc\'{\i}a P\'erez 2001). Full 3D + NLTE calculations
of the UV OH lines are strongly encouraged.

It would be important to check whether \ion{Fe}{1}
and \ion{Fe}{2} lines satisfy the excitation and ionization
equilibrium in 3D+NLTE (and also 1D+NLTE), 
which may bring support (or not) to the
new \teff scale by RM05b. Nevertheless, the NLTE calculations should
be first checked employing stars with well-determined
stellar parameters. This is a formidable theoretical and
observational task, especially for turn-off very metal-poor
stars, where high excitation potential (3-5 eV) \ion{Fe}{1} lines 
are extremely weak, yet important to minimize the
degeneracy of stellar parameters based on
spectroscopic equilibrium of iron lines.
Besides, further work on establishing a reliable
$gf$-scale of \ion{Fe}{1} lines is required.

The determination of a reliable A$_{Fe}^{\rm NLTE}$ is very
important, since the oxygen-to-iron ratio depends
on the adopted Fe abundance. In fact, as discussed in
$\S$1, the relatively low [O/Fe] ratios found in the
reanalysis of UV OH lines by King (2000) was partly due to
large $\delta_{\rm NLTE}$(\ion{Fe}{1}) adopted from 
Th\'evenin \& Idiart (1999).

\subsection{Other flat [O/Fe] ratios and hot \teff scales}

Our finding of a flat (and low) [O/Fe] ratio using the permitted
\ion{O}{1} lines has been previously reported in the literature, 
although in a more restricted metallicity range.
 
This result has been achieved with different \teff scales, 
although it was first shown with the hot \teff scale
by King (1993). Tomkin et al. (1992) obtained a flat
[O/Fe], but with a high mean [O/Fe] $\approx$ +0.8. 
The reanalysis of Tomkin et al. data by King (2000), employing 
new stellar parameters and A$_{Fe}^{\rm NLTE}$,
resulted in a significant reduction of [O/Fe], partly due to high
A$_{FeI}^{\rm NLTE}$ abundances. Unfortunately, King (2000) only
presented his \ion{O}{1} reanalysis in plots, so a detailed comparison 
is not possible.  The [O/Fe] ratios obtained by both 
Tomkin et al. (1992) and King (2000) should be increased 
by about 0.2 dex due to the new low solar O and Fe abundances,
increasing thus the discrepancy of Tomkin et al. 
with the present work.
Note that the NLTE corrections adopted by Tomkin et al. (1992)
are almost negligible, because they 
empirically included the effects of neutral H collisions, 
adjusting them so that the NLTE solar oxygen abundance
from the \ion{O}{1} triplet reproduced the high
solar A$_O$ obtained from the [\ion{O}{1}]
and infrared OH lines (A$_O$ = 8.92). In this way, Tomkin
et al. significantly reduced the departures from LTE.
However, recent studies (see $\S$3.1 for references) show
that the solar O abundance is much lower than previously
thought. Our solar A$_O^{\rm NLTE}$ from the \ion{O}{1} triplet
is in excellent agreement with the O abundance 
from other spectral features (see $\S$3.1),
hence the inclusion of neutral H collisions may not be necessary.

Primas et al. (2001) presented preliminary
NLTE oxygen abundances from the \ion{O}{1} triplet in a sample
of dwarfs with [Fe/H] $\geq -$2.4. Their analysis employed
three different sets of \tsin: the {\it (b-y)} calibrations
of Alonso et al. (1996) and Carney (1983), 
and temperatures from Th\'evenin \& Idiart (1999).
Primas et al. [O/Fe] ratios are roughly flat, 
with [O/Fe] $\approx$ +0.4 for the Alonso et al. (1996) \teff scale,
and [O/Fe] $\approx$ +0.5 for the other two \teff scales.
The preliminary results of Primas et al. are in good agreement with
our work. This is not surprising, since the
{\it (b-y)} calibration of Alonso et al. (1996)
employed by Primas et al. (2001) gives even 
hotter temperatures than those from the RM05b \teff scale,
at least for [Fe/H] $\geq -$2 and \teff $<$ 6000 K (Fig. 7). 
On the other hand, the {\it (b-y)} calibration of
Carney (1983) at [Fe/H] = $-$2 is not much cooler
(only 20 - 100 K lower) than the {\it (b-y)} calibration of RM05b.
It is unknown what \teff scale was employed by
Th\'evenin \& Idiart (1999), since they refereed to
a CDS catalogue by Th\'evenin, where no details are given.

King (1993) found that the very large [O/Fe] ratios 
derived by Abia \& Rebolo (1989) were
partly due to large errors in equivalent widths (overestimated by
about 24\%). Furthermore, as a solution to the discrepancy 
between the oxygen abundances obtained from the triplet and
the forbidden lines, King (1993) suggested an increase of
about 150-200 K in the \teff scale of dwarfs, finding in this
way a flat [O/Fe] $\approx$ 0.5 dex (his Fig. 9).
On the other hand, Carretta et al. (2000), employing the
hot \teff scale of Gratton et al. (1996) and significant 
NLTE corrections by Gratton et al. (1999),
found that the analysis of the permitted oxygen lines results in an
almost flat [O/Fe] $\approx$ 0.5 dex (their Fig. 3), similar
to that obtained from the forbidden oxygen lines.

It is not possible to perform detailed comparisons
with the work of King (1993), since there is only
one star (BD +02 375) in common with our work.
There is a good agreement in both the \teff (ours is only 58 K higher)
and [O/Fe] of that star, but the agreement in [O/Fe] is fortuitous,
because a different [Fe/H] and very small $\delta_{\rm NLTE}$(\ion{O}{1})
were employed by King (1993). 
Larger NLTE corrections would decrease King (1993)
[O/Fe] ratios. On the other hand, 
[O/Fe] should be increased due to the new
low solar A$_O$. Overall, both effects
roughly cancel, preserving thus the [O/Fe] $\approx$ +0.5
found by King (1993).
  
The other well-known example of a flat [O/Fe] is that
of Carretta et al. (2000), who have seven stars in
common with our sample. In Fig. 10 are shown the
differences in [Fe/H], [O/Fe]$_{\rm NLTE}$, A$_O^{\rm LTE}$, $\delta_{\rm NLTE}$(\ion{O}{1})
and \tsin. As can be seen there is a good
agreement in [Fe/H] (Figs. 10a,e) and \teff (Figs. 10d,h). 
Our [O/Fe] ratios are about 0.12 dex 
higher than theirs (Figs 10b,f). This is explained by the much higher
solar O abundance (A$_O$ = 8.93) and the smaller $\delta_{\rm NLTE}$(\ion{O}{1})
adopted by Carretta et al. (2000). After correcting this
(0.28 dex in the solar A$_O$ and 0.12 dex in $\delta_{\rm NLTE}$), 
Carretta et al. (2000) [O/Fe] ratios are only 0.04 dex smaller than ours. 
In fact, an excellent agreement is found in A$_O^{\rm LTE}$ (open circles, Figs. 10c,g),
since it is independent of the solar A$_O$ and $\delta_{\rm NLTE}$.

In summary, despite zero point differences due to the
adopted solar abundances and $\delta_{\rm NLTE}$,
the use of a hot \teff scale and adequate NLTE calculations 
(which reproduce the new solar O abundance) 
result in a flat and low [O/Fe] ratio in halo stars.
Nevertheless, previous claims for a hotter temperature scale have
not been well received by the astronomical community.
The last hot \teff scale recently introduced in the
literature is due to Fulbright \& Johnson (2003), who derived a new 
ad hoc \teff scale for subgiants and giants based on forcing agreement
between the oxygen abundances obtained from permitted and forbidden lines. 
As noted by Fulbright \& Johnson (2003), this hotter \teff scale
for giants does not agree with theoretical 
isochrones\footnote{note that we argue later that the 
hotter IRFM \teff scale of RM05a,b does agree with 
stellar evolution models, but the \teff scale of RM05a,b
is only hotter for metal-poor {\it turn-off} stars,
while the ad hoc \teff scale by Fulbright \& Johnson (2003) is 
hotter for {\it giant} stars},
\teff from Balmer lines, 
and the IRFM \teff scale of Alonso et al. (1996, 1999). 
Interestingly, the ad hoc \teff scale leads to a mild increase
of [O/Fe] for lower metallicities, with [O/Fe] $\approx$ 0.65 at
[Fe/H] = $-1.5$ to [O/Fe] $\approx$ 0.8 at [Fe/H] = $-2.5$.

The new \teff scale by RM05a,b is not based on
ad hoc assumptions on stellar chemical abundances.
It is based on the IRFM, which is probably the least model-dependent
indirect method for determining \tsin. The problem with
previous hot \teff scales is that they basically adopted
a zero point shift, that is, the increase in effective
temperatures was applied to all stars independent of
spectral type, which is not correct. Our recent studies
(RM05a,b) have shown that for most of the range
spanned by stellar atmosphere parameters of FGK stars 
the IRFM \teff scale of Alonso et al. (1996, 1999) remains roughly valid,
except for some regions where insufficient data was previously
available, which is the case of metal-poor F and early G dwarfs.

Besides its impact on the oxygen abundances derived from the 
\ion{O}{1} triplet, the new IRFM \teff scale (RM05a,b)
may also affect the abundances derived from other high excitation
lines (e.g. \ion{C}{1}, \ion{S}{1}), as well as low excitation lines 
(e.g. \ion{Li}{1}). Indeed, Mel\'endez \& Ram\'{\i}rez (2004) have shown
that the use of the new IRFM \teff scale may be one of the
factors concurring to alleviate the 
discrepancy between the ``low'' Li abundance obtained in 
metal-poor FG dwarfs and the ``high'' primordial Li abundance 
obtained from WMAP data and Big Bang primordial nucleosynthesis.

\subsection{New IRFM \teff scale for halo turn-off stars vs. Stellar Evolution}

The high effective temperatures that we obtained for metal-poor turn-off stars 
are supported by stellar evolution calculations, which predict
that the \teff of the turn-off strongly depends on metallicity.
In Fig. 11, we plot Y$^2$ isochrones for metallicities [Fe/H] = $-$1, $-$2, $-$3,
showing that the turn-off temperature
increases up to $\approx$ 700 K from [Fe/H] = $-$1 to [Fe/H] = $-$3,
assuming a constant age of 12 Gyrs. Even allowing for reasonable
changes in age, the \teff of the turn-off still strongly depends on metallicity.
This metallicity dependence of the turn-off temperatures are also predicted
by other stellar evolution models. The widely used Padova 
(e.g. Girardi et al. 2002)
and Victoria (Bergbusch \& VandenBerg 2001) isochrones 
predict a strong dependence on metallicity, with even higher effective
temperatures at the turn-off than the Y$^2$ isochrones.\footnote{The absolute location 
of the turn-off depends on the assumptions made on the stellar evolution models,
especially on the mixing length. However, our main
point is that the turn-off strongly depends on metallicity,
and this well established behavior of stellar evolution
is independent of the absolute location of the turn off.}

In order to study more quantitatively whether (or not) our 
IRFM \teff scale for turn-off stars makes sense physically, we have made
a comparison of our temperatures
with those predicted by stellar evolution.
Since for a given metallicity the turn-off stars are the ones 
with highest \tsin, we divided the present sample and 
the sample of Mel\'endez \& Ram\'{\i}rez (2004) (which are
composed mainly of main sequence, turn-off and subgiant stars)
in metallicity bins of 0.5 dex, and adopted as turn-off 
IRFM \teff the average effective temperature of the three stars 
with the highest \teff in each metallicity bin.
The results are shown in Fig. 12, where we can see the 
impressive qualitative and quantitative agreement of the 
evolutionary models of metal-poor stars with the \teff scale by RM05a,b.

The metallicity dependence of the turn-off was predicted by
early stellar evolution models 
(Simoda \& Iben 1968, 1970; Iben \& Rood 1970; 
Demarque, Mengel \& Aizenman 1971), 
which showed that a decrease by a factor of 10 in metallicity 
(from Z = 10$^{-3}$  to 10$^{-4}$) corresponds 
to an increase of about 0.025 dex in log \teff
(Iben \& Rood 1970), that at \teff = 6000 K corresponds to an increase in
\teff of 355 K for a decrease in 1 dex in metallicity. 
This is due to two effects (Simoda \& Iben 1968, 1970):
{\it i)} the influence of metallicity on opacity,
which contributes to an increase of the \teff for 
more metal-poor models; and {\it ii)} the influence of 
metallicity on the $p$-$p$ and CNO luminosities
for low mass stars. As evolution from the main sequence
proceeds and core temperature increases, the CN-cycle
grows its importance over the $p$-$p$ chain. 
The CN-cycle does not become important until a
considerable fraction of the central hydrogen has been
converted into helium, which for lower metallicities occurs later, 
allowing the star to reach a bluer (hotter) turn-off
(Iben \& Rood 1970; Simoda \& Iben 1970). 


Recently, Frebel et al. (2005) announced the discovery
of HE 1327-2326, a hyper metal-poor star with [Fe/H] = $-$5.4,
the most iron-poor star yet known.
The authors argue that this star is close to the
turn-off, being probably either a main-sequence or
a subgiant star. 

Models of hyper metal-poor stars show
that the location of the turn-off looses its sensitivity
to the initial metal content for
$Z$ lower than $Z \approx 10^{-6}$
(Wagner 1974; Cassisi \& Castellani 1993).
That means that the \teff of the turn-off
reaches a maximum value at [Fe/H] $\approx -4$.
Fig. 12 suggests that the turn-off \teff of 
population III stars should be not much higher
than $\approx$ 6700-6750 K. 

Frebel et al. (2005) determined a \teff = 6180 $\pm$ 80 K for HE 1327-2326,
which is considerably lower than the maximum temperature
allowed for a hyper metal-poor turn-off star. 
Employing the Johnson-Cousin and 2MASS colors given
in Aoki et al. (2005), and using E(B-V) = 0.077, 
which is the same E(B-V) value adopted by them, 
we obtained \teff = 6340 K using the 
\teff scale by RM05b with [Fe/H] = $-$3.3. A lower
metallicity must not be employed due to the 
low number of extremely metal-poor calibration stars.
Aoki et al. (2005) found that the reddening of
HE 1327-2326 may be as high as E(B-V) = 0.104, as
estimated from the interstellar \ion{Na}{1} D2 line.
Adopting E(B-V) = 0.104, \teff = 6500 K is obtained
from the \teff scale of RM05b. 
Aoki et al. (2005) have found that the spectroscopic
temperature (based on Balmer lines and the HP2 index) 
of HE 1327-2326 is about 200-300 K lower than G 64-12.
Using our \teff for G 64-12, implies that the
spectroscopic \teff of HE 1327-2326 is about 6360-6460 K.

It is of the uttermost importance to determine
a good parallax for HE 1327-2326, in order to assess the 
evolutionary stage of this hyper metal-poor star,
and to further constraint its \tsin. For example, 
a trigonometric gravity of log $g$ = 4.1 $\pm$ 0.1 dex
would constraint HE 1327-2326 to the turn-off (within $\approx$ 200 K),
and if this star turns out to be hotter than previously
thought, then the obtained abundances and their 
interpretation might require revision. This is
especially important for the abundance of elements very
sensitive to \tsin, like the oxygen abundance determined
from UV OH lines.

\section{Conclusions}

We have determined oxygen and iron abundances in
31 metal-poor ($-3.2 <$ [Fe/H] $< -0.7$) stars 
close to the turn-off, employing 
high resolution high S/N UVES data taken from Ake04 and N04.

We find a flat [O/Fe] = +0.5, independent of metallicity,
temperature and surface gravity in the ranges $-3.2 <$ [Fe/H] $< -0.7$,
5700 K $<$ \teff $<$ 6700 K, and $3.7 <$ log $g$ $< 4.5$, respectively.
Our work confirms previous studies (e.g. Carretta et al. 2000; Primas et al. 2001)
which have already found a flat and low [O/Fe] ratio in halo dwarfs from the \ion{O}{1} triplet,
extending the constancy of the [O/Fe] ratios down to [Fe/H] = $-$3.2.

The flat [O/Fe] ratio is mainly due to the use of 
adequate NLTE calculations 
and the new IRFM \teff scale by RM05a,b, which for metal-poor
turn-off stars is hotter than most previous \teff scales
available in the literature.
We find a low star-to-star scatter of 0.136 dex for
the whole sample, or $\sigma$ = 0.10 dex for the
sample with low reddening uncertainty. 

The observed star-to-star scatter ($\sigma_{obs}$ = 0.136 dex) 
is almost completely explained by errors in the analysis
($\approx$ 0.121 dex), leaving little room for intrinsic scatter.
Hence, the Galaxy was extremely efficient in 
mixing the chemical elements ejected by supernovae.
Other recent works in the literature have also found very small 
star-to-star scatter for other $\alpha$-elements 
(e.g. Cayrel et al. 2004; Cohen et al. 2004; Arnone et al. 2005).
Furthermore, the low scatter implies a small contribution
to the halo from metal-poor stars that originated in
dSph galaxies, since much lower [$\alpha$/Fe] ratios
are commonly seen in such galaxies (e.g. Venn et al. 2004;
see also discussion and references given in Catelan 2005).

The constancy of the [O/Fe] ratio
over more than two orders of magnitude in [Fe/H], 
from [Fe/H] = $-$0.7 to [Fe/H] = $-$3.2, 
is telling us that the formation
of the halo was extremely fast, with a timescale shorter
than the bulk of Type Ia SNe. Our data provides tight constraints
for Galactic chemical evolution models (e.g. Matteucci \& Recchi 2001; 
Alib\'es, Labay \& Canal 2001; Goswami \& Prantzos 2000;
Portinari, Chiosi \& Bressan 1998;
Samland 1998; Chiappini, Matteuci \& Gratton 1997).

Our low [O/Fe] = +0.5 constrains the $^6$Li production
by GCR models, which are not able to explain the
detection of $^6$Li in a star with [Fe/H] = $-$2.7 (Asplund et al. 2005),
hence reinforcing the case for a pre-Galactic origin of the
recent $^6$Li detections in very metal-poor stars.

Recent determinations of H$\alpha$ temperatures 
in turn-off stars by Asplund et al. (2005)
are in very good relative agreeement with the 
RM05b \teff scale in the metallicity range $-2.7 <$ [Fe/H] $< -1$,
with a zero point difference of only 61$\pm$62 K.

As shown in Figs. 11 and 12, the strong metallicity dependence of the 
temperature of turn-off stars is not unphysical, but a 
natural consequence of stellar evolution. Ram\'{\i}rez et al. (2006),
has recently confirmed the hot \teff scale of RM05b for one star, BD +17 4708,
by fitting its observed absolute flux distribution from
Hubble Space Telescope observations with Kurucz models. Indeed, the
\teff obtained by Ram\'{\i}rez is about 100 K higher than the \teff obtained
in this work with the \teff scale of RM05b. This is not in conflict
with stellar evolution calculations, since there is still room for
an increase of about 100 K in the RM05b \teff scale of metal-poor
turn-off stars (Fig. 12).

In the future, it would be important to take into account also
granulation effects, performing full 3D+NLTE analyses.
This is the way to go for future abundance studies.
Also, it is very important to
perform future verifications of the \teff scale of RM05b.

\acknowledgements
J.M. thanks A. McWilliam and C. Allende Prieto for
providing Fortran and IDL codes to interpolate model atmospheres,
thanks partial support from NSF grant AST-0205951 to J. G. Cohen,
and acknowledges the support of the American Astronomical Society 
and the NSF in the form of two International Travel Grants.
I.R. acknowledges support from the Robert A. Welch Foundation
of Houston, Texas to D. Lambert. 
We thank an anonymous referee for useful suggestions;
K. A. Olive \& B. D. Fields for sending in electronic form data plotted 
in Fields et al. (2005); M. Catelan for his comments, especially
on stellar evolution; C. Chiappini for comments on early results;
J. G. Cohen and C. Dickinson for
comments and proofreading of an early version of the manuscript; 
M. Asplund for useful comments, especially for 
confirming errors in the NLTE corrections of the most metal-poor stars in Ake04;
A. E. Garc\'{\i}a-P\'erez for sending a pre-print prior to publication;
P. Nissen for clarifying the solar A$_{Fe}$ of Nissen et al. (2002) and N04;
and D. Fabbian for useful discussions on NLTE effects.
This research has been funded partially by the Spanish Ministerio de Educaci\'on
y Ciencia through project AYA2004-05792 and by the European Commission through
INTAS grant 00-00084.
We  have made use of data from: the Hipparcos astrometric mission of the ESA;
UVES/VLT of ESO; SIMBAD database operated at CDS; and 2MASS of
the University of Massachusetts and IPAC/Caltech, funded by NASA
and NSF.

\clearpage

\begin{deluxetable}{llllll}
\tablewidth{0pt}
\tablecolumns{6}
\scriptsize
\tablecaption{\ion{O}{1} triplet}
\tablehead{
\colhead{$\lambda$}   &
\colhead{$\chi_{exc}$} &
\colhead{log {\it gf}} &
\multicolumn{3}{c}{$W_\lambda${$_\odot$}} \\ 
\colhead{} &
\colhead{} &
\colhead{} &
\colhead{K84} &
\colhead{UVES} &
\colhead{Asp04} \\
\colhead{({\rm \AA})}   &
\colhead{(eV)}  &
\colhead{(dex)} &
\multicolumn{3}{c}{({\rm m\AA})}
}
\startdata
 7771.944  & 9.1461 &0.37 &  71.5 & 72.0& 71.2 \\
 7774.166  & 9.1461 &0.22 &  61.9 & 61.8& 61.8 \\
 7775.388  & 9.1461 &0.00 &  48.6 & 47.4& 48.8 \\
\enddata
\label{triplet}
\end{deluxetable}

\begin{deluxetable}{lllll}
\tablewidth{0pt}
\tablecolumns{5}
\scriptsize
\tablecaption{\ion{Fe}{2} lines}
\tablehead{
\colhead{$\lambda$}   &
\colhead{$\chi_{exc}$} &
\colhead{log {\it gf}} &
\multicolumn{2}{c}{$W_\lambda${$_\odot$}} \\ 
\colhead{} &
\colhead{} &
\colhead{} &
\colhead{K84}  &
\colhead{UVES} \\
\colhead{({\rm \AA})}   &
\colhead{(eV)}  &
\colhead{(dex)} &
\multicolumn{2}{c}{({\rm m\AA})}
}
\startdata
 4128.748$^\P$  & 2.5828 &$-$3.63 & & \\
 4178.862$^\P$  & 2.5828 &$-$2.51 & & \\
 4233.172$^\P$  & 2.5828 &$-$1.97 & & \\
 4413.601      & 2.6759 &$-$3.79 & 38.3 & 39.5\\
 4416.830$^\P$  & 2.7786 &$-$2.65 & & \\
 4489.183$^\P$  & 2.8283 &$-$2.96 & & \\
 4491.405$^\P$  & 2.8557 &$-$2.71 & 69.0 & 70.6 \\
 4508.288$^\P$  & 2.8557 &$-$2.44 & 83.5 & 82.9 \\
 4515.339$^\P$  & 2.8443 &$-$2.60 & & \\
 4520.224$^\P$  & 2.8068 &$-$2.65 & 75.0 & 76.1 \\
 4522.634$^\P$  & 2.8443 &$-$2.25 & & \\
 4534.168      & 2.8557 &$-$3.28 & 51.3 & 51.9 \\
 4541.524$^\P$  & 2.8557 &$-$2.98 & 61.0 & 60.6 \\
 4555.893$^\P$  & 2.8283 &$-$2.40 & 86.5 & 84.3 \\
 4576.340$^\P$* & 2.8443 &$-$2.95 & 63.5 & 62.5 \\
 4582.835$^\P$  & 2.8443 &$-$3.18 & 52.9 & 52.9 \\
 4583.837$^\P$  & 2.8068 &$-$1.93 & & \\
 4620.521$^\P$* & 2.8283 &$-$3.21 & 49.5 & 51.2 \\
 4656.981$^\P$* & 2.8912 &$-$3.60 & 33.1 & 31.1 \\
 4666.758$^\P$  & 2.8283 &$-$3.28 & 47.0 & 48.6 \\ 
 4923.927$^\P$  & 2.8912 &$-$1.26 & & \\
 5197.577      & 3.2306 &$-$2.22 & 79.9 & 81.8 \\
 5234.625*     & 3.2215 &$-$2.18 & 81.0 & 83.2 \\
 5264.812*     & 3.2304 &$-$3.13 & 46.1 & 45.4 \\
 5325.553      & 3.2215 &$-$3.16 & 42.1 & 41.2 \\
 5414.073*     & 3.2215 &$-$3.58 & 26.5 & 25.6 \\
 5425.257      & 3.1996 &$-$3.22 & 40.7 & 41.6 \\
 6369.462      & 2.8912 &$-$4.11 & 18.7 & 18.3 \\
 6432.680*     & 2.8912 &$-$3.57 & 41.3 & 40.7 \\
 6516.080*     & 2.8912 &$-$3.31 & 53.3 & 52.5 \\
 7222.394*     & 3.8889 &$-$3.26 & 19.4 & 18.5 \\
 7224.487*     & 3.8891 &$-$3.20 & 18.9 & 18.9 \\
 7449.335*     & 3.8889 &$-$3.27 & 17.1 & 18.2 \\
 7515.832*     & 3.9036 &$-$3.39 & 13.3 & 13.5 \\
 7711.724*     & 3.9034 &$-$2.50 & 45.9 & 46.7 \\
\enddata
\tablecomments{($^\P$) line used for halo stars \\ 
(*) line in common with H92 (see Fig. 1).}
\label{fe2}
\end{deluxetable}

\begin{deluxetable}{llllll}
\tablewidth{0pt}
\tablecolumns{6}
\scriptsize
\tablecaption{Solar Fe Abundance and $\sigma$* from
different sets of \ion{Fe}{2} lines**}
\tablehead{
\colhead{Model}   &
\colhead{{\it gf}-values} &
\colhead{{\it gf}-values} &
\colhead{{\it gf}-values$^\P$} \\
\colhead{atmosphere}   &
\colhead{(H92)} &
\colhead{(this work)} &
\colhead{(GK89)}
}
\startdata
Kurucz  & 7.46 (0.11) &  7.45 (0.05) & 7.63 (0.05) \\
MARCS   & 7.43 (0.11) &  7.42 (0.06) & 7.60 (0.06) \\
$<$3D$>$ & 7.44 (0.11) &  7.43 (0.05) & 7.61 (0.06) \\
HM74    & 7.46 (0.11) &  7.45 (0.05) & 7.63 (0.04) \\
\enddata
\tablecomments{(*) $\sigma$ is given in parenthesis\\
(**) We used 13 \ion{Fe}{2} lines in common 
between this work (Table 2), H92 and GK89 \\
($^\P$) solar oscillator strengths}
\label{sigmaFe}
\end{deluxetable}

\clearpage
\thispagestyle{empty}
\begin{deluxetable}{lllllllllllll}
\tablewidth{0pt}
\tabletypesize{\footnotesize}
\tablecaption{Reddening, Atmospheric Parameters and Chemical Abundances$^\P$}
\tablehead{
\colhead{ID} & \colhead{$E_{B-V}$} & \colhead{\teff} &   
\colhead{log $g$} & \colhead{[M/H]} & \colhead{v$_t$} &
 \multicolumn{3}{c}{\ion{Fe}{2}} & \multicolumn{3}{c}{\ion{O}{1}} & \colhead{[O/Fe]}\\
\colhead{}  & \colhead{}  & \colhead{} &
\colhead{}  & \colhead{} & \colhead{} &
\colhead{A$_{\rm LTE}$} & \colhead{A$_{\rm NLTE}$} & \colhead{[Fe/H]$_{\rm NLTE}^{**}$} &
\colhead{A$_{\rm LTE}$} & \colhead{A$_{\rm NLTE}$} & \colhead{[O/H]$_{\rm NLTE}^{**}$} &
\colhead{} \\
\colhead{}  & \colhead{(mag)}  & \colhead{(K)} &
\colhead{(dex)}  & \colhead{(dex)} & \colhead{km s$^{-1}$} &
\colhead{(dex)} & \colhead{(dex)} & \colhead{(dex)} &
\colhead{(dex)} & \colhead{(dex)} & \colhead{(dex)} &
\colhead{(dex)} }
\startdata
BD+023375  & 0.022$\pm$0.022 & 6045$\pm$ 69 & 4.18$\pm$0.12 &-2.0 & 1.5 & 5.26 & 5.32 & $-$2.13$\pm$0.08& 7.43 & 7.25 & $-$1.40$\pm$0.12 & 0.73$\pm$0.18\\
BD+024651  & 0.019$\pm$0.007 & 6132$\pm$ 58 & 3.85$\pm$0.15 &-1.6 & 1.8 & 5.71 & 5.76 & $-$1.69$\pm$0.07& 7.74 & 7.53 & $-$1.12$\pm$0.02 & 0.57$\pm$0.09\\
BD+083095  & 0.012$\pm$0.012 & 5657$\pm$ 41 & 4.21$\pm$0.30 &-0.6 & 1.3 & 6.72 & 6.75 & $-$0.70$\pm$0.10& 8.75 & 8.48 & $-$0.17$\pm$0.02 & 0.53$\pm$0.13\\
BD+174708  & 0.009$\pm$0.008 & 6091$\pm$ 92 & 4.01$\pm$0.16 &-1.4 & 1.8 & 5.86 & 5.92 & $-$1.53$\pm$0.08& 7.95 & 7.72 & $-$0.93$\pm$0.05 & 0.60$\pm$0.13\\
BD-043208  & 0.013$\pm$0.013 & 6376$\pm$ 59 & 3.84$\pm$0.15 &-2.1 & 1.5 & 5.17 & 5.20 & $-$2.25$\pm$0.07& 7.17 & 6.98 & $-$1.67$\pm$0.03 & 0.58$\pm$0.11\\
BD-133442  & 0.018$\pm$0.021 & 6510$\pm$ 40 & 4.10$\pm$0.30 &-2.5 & 1.5 & 4.89 & 4.92 & $-$2.53$\pm$0.06& 6.89 & 6.67 & $-$1.98$\pm$0.07 & 0.55$\pm$0.13\\
CD-3018140 & 0.023$\pm$0.004 & 6305$\pm$ 44 & 4.28$\pm$0.12 &-1.7 & 1.9 & 5.63 & 5.68 & $-$1.77$\pm$0.08& 7.66 & 7.46 & $-$1.19$\pm$0.04 & 0.58$\pm$0.10\\
CD-3514849 & 0.016$\pm$0.003 & 6313$\pm$ 43 & 4.33$\pm$0.15 &-2.0 & 1.5 & 5.17 & 5.21 & $-$2.24$\pm$0.07& 7.11 & 6.95 & $-$1.70$\pm$0.06 & 0.54$\pm$0.10\\
CD-4214278 & 0.020$\pm$0.009 & 6006$\pm$ 42 & 4.52$\pm$0.15 &-1.8 & 1.4 & 5.51 & 5.59 & $-$1.86$\pm$0.07& 7.41 & 7.25 & $-$1.40$\pm$0.02 & 0.46$\pm$0.09\\
G011-044   & 0.005$\pm$0.005 & 6092$\pm$ 47 & 4.49$\pm$0.15 &-1.9 & 1.5 & 5.49 & 5.56 & $-$1.89$\pm$0.07& 7.52 & 7.35 & $-$1.30$\pm$0.04 & 0.59$\pm$0.09\\
G024-003   & 0.039$\pm$0.025 & 6077$\pm$ 35 & 4.41$\pm$0.15 &-1.4 & 1.6 & 5.92 & 5.98 & $-$1.47$\pm$0.08& 7.67 & 7.49 & $-$1.16$\pm$0.06 & 0.31$\pm$0.15\\
G053-041   & 0.017$\pm$0.015 & 5970$\pm$ 27 & 4.38$\pm$0.15 &-1.1 & 1.4 & 6.24 & 6.29 & $-$1.16$\pm$0.08& 7.89 & 7.71 & $-$0.94$\pm$0.02 & 0.22$\pm$0.11\\
G064-012   & 0.017$\pm$0.019 & 6660$\pm$ 77 & 4.24$\pm$0.15 &-3.0 & 1.5 & 4.26 & 4.28 & $-$3.17$\pm$0.10& 6.36 & 6.01 & $-$2.64$\pm$0.11 & 0.53$\pm$0.18\\
G064-037   & 0.010$\pm$0.001 & 6658$\pm$ 61 & 4.27$\pm$0.15 &-2.8 & 1.5 & 4.44 & 4.46 & $-$2.99$\pm$0.06& 6.35 & 6.02 & $-$2.63$\pm$0.04 & 0.36$\pm$0.09\\
G066-030   & 0.016$\pm$0.007 & 6316$\pm$ 38 & 4.18$\pm$0.15 &-1.3 & 1.8 & 5.91 & 5.95 & $-$1.50$\pm$0.08& 7.99 & 7.74 & $-$0.91$\pm$0.02 & 0.59$\pm$0.10\\
HD103723   & 0.020$\pm$0.013 & 6013$\pm$ 21 & 4.26$\pm$0.12 &-0.6 & 1.4 & 6.66 & 6.69 & $-$0.76$\pm$0.10& 8.41 & 8.14 & $-$0.51$\pm$0.02 & 0.25$\pm$0.12\\
HD106038   & 0.007$\pm$0.006 & 6012$\pm$ 26 & 4.45$\pm$0.11 &-1.2 & 1.7 & 6.08 & 6.14 & $-$1.31$\pm$0.08& 8.11 & 7.90 & $-$0.75$\pm$0.03 & 0.56$\pm$0.10\\
HD108177   & 0.003$\pm$0.005 & 6133$\pm$ 49 & 4.42$\pm$0.1* &-1.5 & 1.8 & 5.81 & 5.86 & $-$1.59$\pm$0.08& 7.85 & 7.65 & $-$1.00$\pm$0.04 & 0.59$\pm$0.10\\
HD110621   & 0.013$\pm$0.004 & 6083$\pm$ 51 & 4.01$\pm$0.12 &-1.4 & 1.7 & 5.84 & 5.88 & $-$1.57$\pm$0.08& 7.95 & 7.73 & $-$0.92$\pm$0.08 & 0.65$\pm$0.12\\
HD140283   & 0.007$\pm$0.011 & 5753$\pm$ 61 & 3.70$\pm$0.1* &-2.2 & 1.4 & 5.12 & 5.20 & $-$2.25$\pm$0.07& 7.13 & 6.97 & $-$1.68$\pm$0.03 & 0.57$\pm$0.10\\
HD146296   & 0.009$\pm$0.011 & 5733$\pm$ 30 & 4.13$\pm$0.11 &-0.6 & 1.4 & 6.69 & 6.73 & $-$0.72$\pm$0.12& 8.49 & 8.24 & $-$0.41$\pm$0.03 & 0.31$\pm$0.14\\
HD148816   & 0.002$\pm$0.002 & 5825$\pm$ 38 & 4.14$\pm$0.1* &-0.6 & 1.5 & 6.68 & 6.71 & $-$0.74$\pm$0.11& 8.67 & 8.38 & $-$0.27$\pm$0.02 & 0.47$\pm$0.12\\
HD160617   & 0.022$\pm$0.005 & 6065$\pm$ 37 & 3.82$\pm$0.10 &-1.6 & 1.8 & 5.70 & 5.75 & $-$1.70$\pm$0.08& 7.43 & 7.25 & $-$1.40$\pm$0.05 & 0.30$\pm$0.10\\
HD179626   & 0.018$\pm$0.002 & 5818$\pm$ 56 & 3.85$\pm$0.11 &-1.0 & 1.5 & 6.28 & 6.33 & $-$1.12$\pm$0.10& 8.40 & 8.13 & $-$0.52$\pm$0.05 & 0.60$\pm$0.13\\
HD181743   & 0.007$\pm$0.007 & 6021$\pm$ 80 & 4.45$\pm$0.11 &-1.6 & 1.5 & 5.64 & 5.70 & $-$1.75$\pm$0.08& 7.65 & 7.47 & $-$1.18$\pm$0.02 & 0.57$\pm$0.11\\
HD188031   & 0.016$\pm$0.016 & 6196$\pm$ 63 & 4.13$\pm$0.30 &-1.5 & 1.5 & 5.77 & 5.81 & $-$1.64$\pm$0.06& 7.77 & 7.55 & $-$1.10$\pm$0.04 & 0.54$\pm$0.11\\
HD193901   & 0.002$\pm$0.002 & 5721$\pm$ 38 & 4.52$\pm$0.1* &-0.9 & 1.2 & 6.41 & 6.46 & $-$0.99$\pm$0.09& 8.21 & 8.01 & $-$0.64$\pm$0.02 & 0.35$\pm$0.11\\
HD194598   & 0.000$\pm$0.000 & 5970$\pm$ 56 & 4.31$\pm$0.1* &-0.9 & 1.5 & 6.32 & 6.37 & $-$1.08$\pm$0.07& 8.22 & 7.99 & $-$0.66$\pm$0.05 & 0.42$\pm$0.11\\
HD215801   & 0.001$\pm$0.001 & 6085$\pm$ 34 & 3.83$\pm$0.15 &-2.0 & 1.5 & 5.23 & 5.29 & $-$2.16$\pm$0.08& 7.28 & 7.10 & $-$1.55$\pm$0.04 & 0.61$\pm$0.10\\
HD340279   & 0.020$\pm$0.012 & 6521$\pm$102 & 4.31$\pm$0.15 &-2.3 & 1.5 & 4.94 & 4.97 & $-$2.48$\pm$0.08& 6.69 & 6.49 & $-$2.16$\pm$0.12 & 0.32$\pm$0.17\\
LP0815-43  & 0.024$\pm$0.015 & 6622$\pm$ 55 & 4.28$\pm$0.15 &-2.4 & 1.5 & 4.85 & 4.87 & $-$2.58$\pm$0.06& 6.58 & 6.34 & $-$2.31$\pm$0.21 & 0.27$\pm$0.23\\ 
\enddata  	   
\tablenotetext{$^\P$}{A(Fe)$_\odot$ = 7.45 and A(O)$_\odot$ = 8.65 have been adopted}
\tablenotetext{*}{a minimum error of 0.1 dex was has been adopted; the calculated error is smaller}
\tablenotetext{**}{the error given here is only the line-to-line scatter}
\label{tab4}   
\end{deluxetable}
\clearpage

\begin{deluxetable}{lrrrr}
\tablewidth{0pt}
\tablecolumns{5}
\tabletypesize
\footnotesize
\tablecaption{Sensitivity to  \tsin, log {\it g}, v$_t$ and [M/H]}
\tablehead{
\colhead{Abundance}  &
\colhead{$\Delta$ \teff} &
\colhead{$\Delta$ log g} &
\colhead{$\Delta$ v$_t$} &
\colhead{$\Delta$ [M/H]} \\
\colhead{}  &
\colhead{+50 K} &
\colhead{+0.15 dex} &
\colhead{+0.3 km s$^{-1}$} &
\colhead{+0.2 dex} \\
}
\startdata
\multicolumn{5}{c}{BD-133442 (6510/4.10/-2.53)*} \\
\hline\\
{[FeII/H]} &   +0.01 &  +0.05 & $-$0.01 &   0.00 \\
{[OI/H]}    & $-$0.03 & +0.06 &    0.00 &   0.00 \\
{[OI/FeII]} & $-$0.04 & +0.01 &   +0.01 &   0.00 \\
\cutinhead{BD+023375 (6045/4.18/-2.13)*}
{[FeII/H]} &   +0.01 &  +0.05 & $-$0.01 &   0.01 \\
{[OI/H]}    & $-$0.03 & +0.05 &    0.00 & $-$0.01 \\
{[OI/FeII]} & $-$0.04 & +0.00 &   +0.01 & $-$0.02 \\
\cutinhead{HD140283 (5753/3.70/-2.25)*} 
{[FeII/H]} &   +0.01 &  +0.05 & $-$0.02 &   0.00 \\
{[OI/H]}    & $-$0.03 & +0.05 &    0.00 &   0.00 \\
{[OI/FeII]} & $-$0.04 &  0.00 &   +0.02 &   0.00 \\
\cutinhead{HD160617 (6065/3.82/-1.70)*}
{[FeII/H]} &   +0.00 &  +0.04 & $-$0.03 &   0.00 \\
{[OI/H]}    & $-$0.03 & +0.04 & $-$0.01 & $-$0.01 \\
{[OI/FeII]} & $-$0.03 &  0.00 &   +0.02 & $-$0.01 \\
\cutinhead{G 53-41 (5970/4.38/-1.16)*}
{[FeII/H]} &    0.00 &  +0.05  & $-$0.05 &  +0.01 \\
{[OI/H]}    & $-$0.04 & +0.04  & $-$0.01 & $-$0.01 \\
{[OI/FeII]} & $-$0.04 & $-$0.01 & +0.04  & $-$0.02 \\
\enddata
\tablenotetext{*}{(\tsin/log {\it g}/[Fe/H])}
\label{errors}
\end{deluxetable}

\clearpage

\begin{figure}
\epsscale{}
\plotone{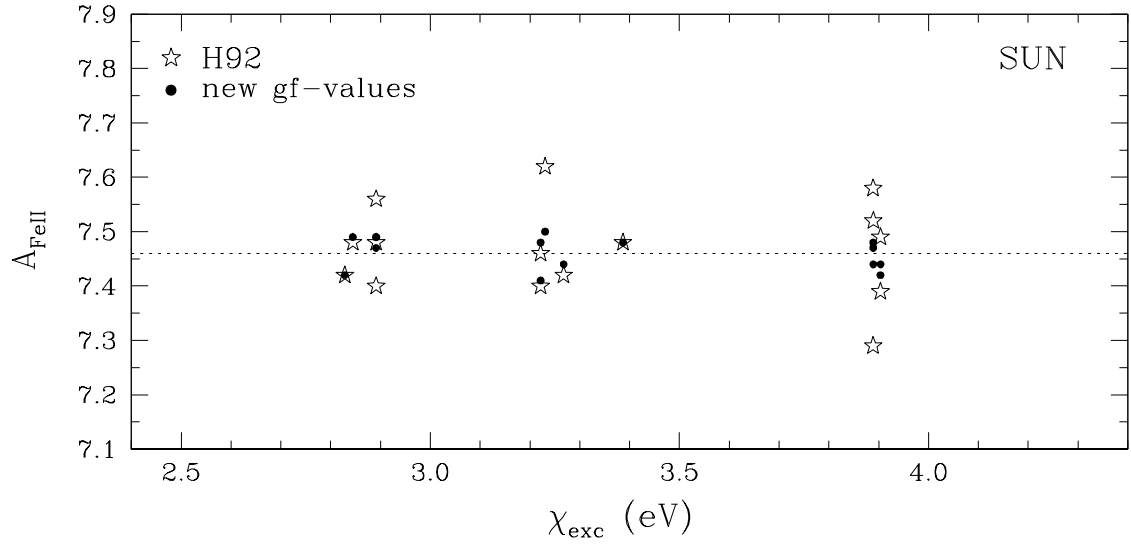}
\caption{Solar iron abundances from \ion{Fe}{2} lines vs. 
excitation potential (H92, stars), 
and rescaled abundances employing our new $gf$-values (filled
circles). The scatter has been significantly reduced due 
to our improved oscillator strengths}
\label{H92} 
\end{figure}

\begin{figure}
\epsscale{}
\plotone{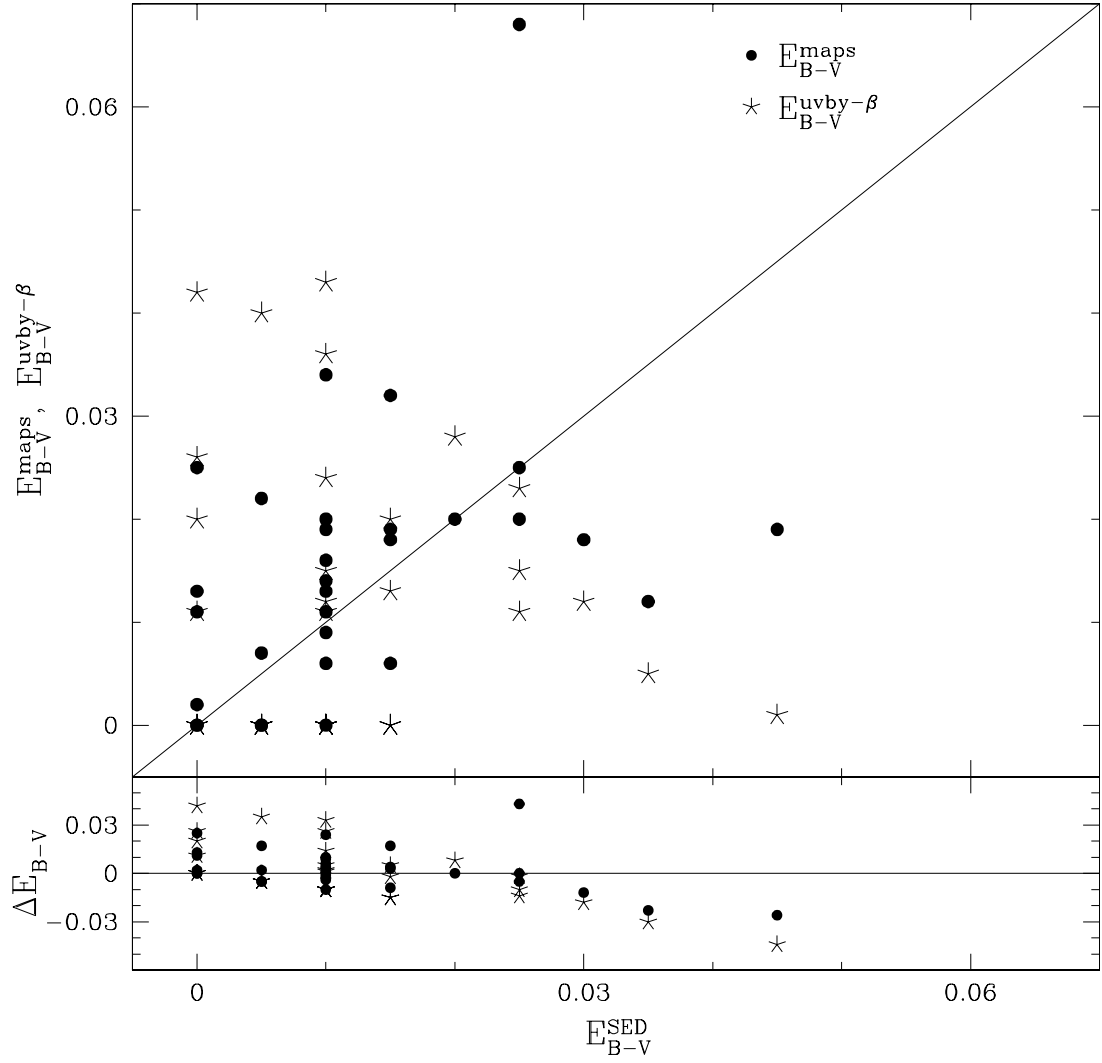}
\caption{Comparison between extinction obtained from maps ($E_{B-V}^{\rm maps}$), 
Str\"omgren photometry ($E_{B-V}^{\rm uvby-\beta}$),
and Stellar Energy Distributions ($E_{B-V}^{\rm SED}$).}
\label{ebv}
\end{figure}

\begin{figure}
\epsscale{}
\plotone{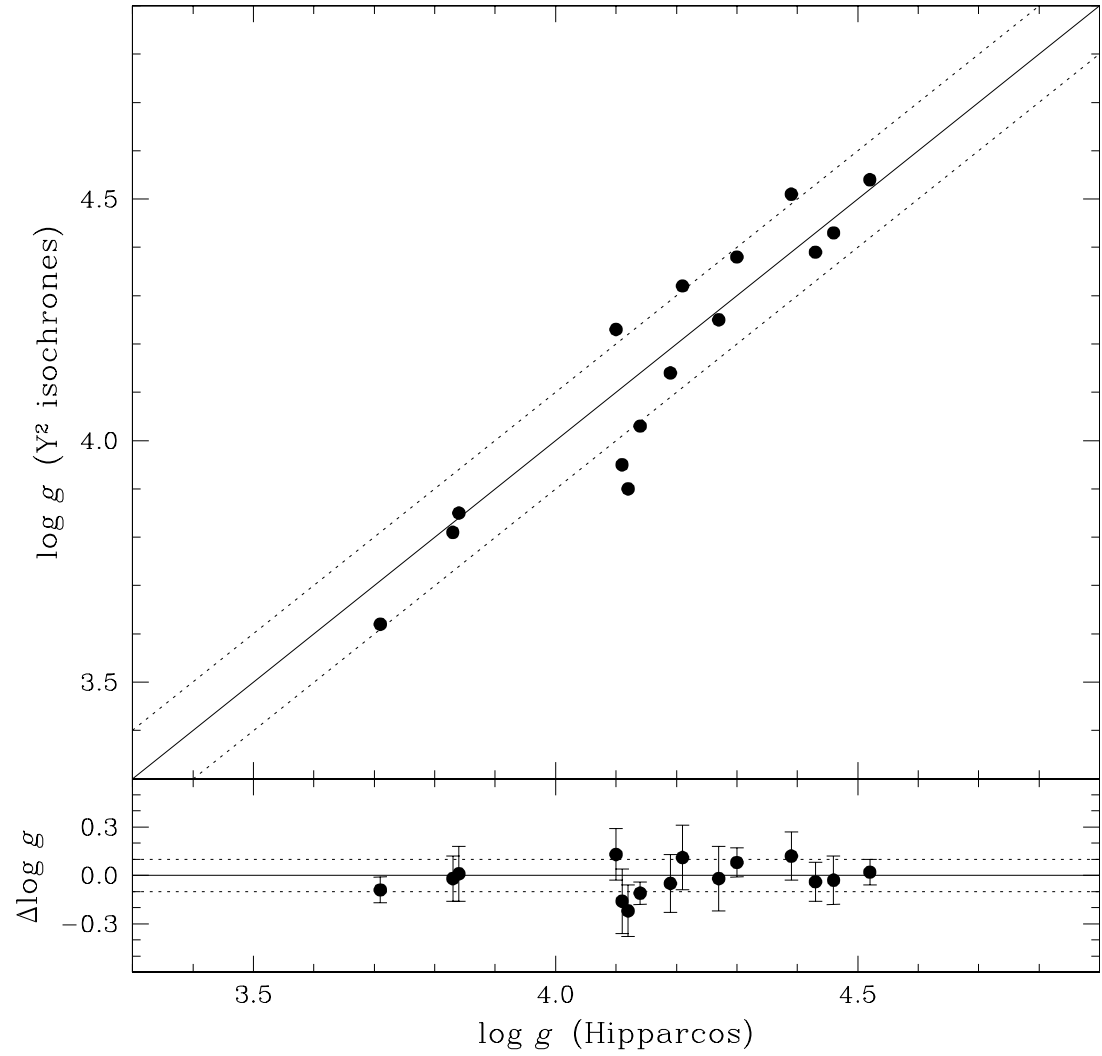}
\caption{Comparison of surface gravities determined from
Hipparcos parallaxes and Y$^2$ isochrones. The error bars
include 1-$\sigma$ errors in Hipparcos parallaxes and 
errors in mass and temperature. Solid and dotted lines 
correspond to the 1:1 line and $\pm$0.1 dex uncertainties, respectively.}
\label{logg}
\end{figure}

\begin{figure}
\epsscale{}
\plotone{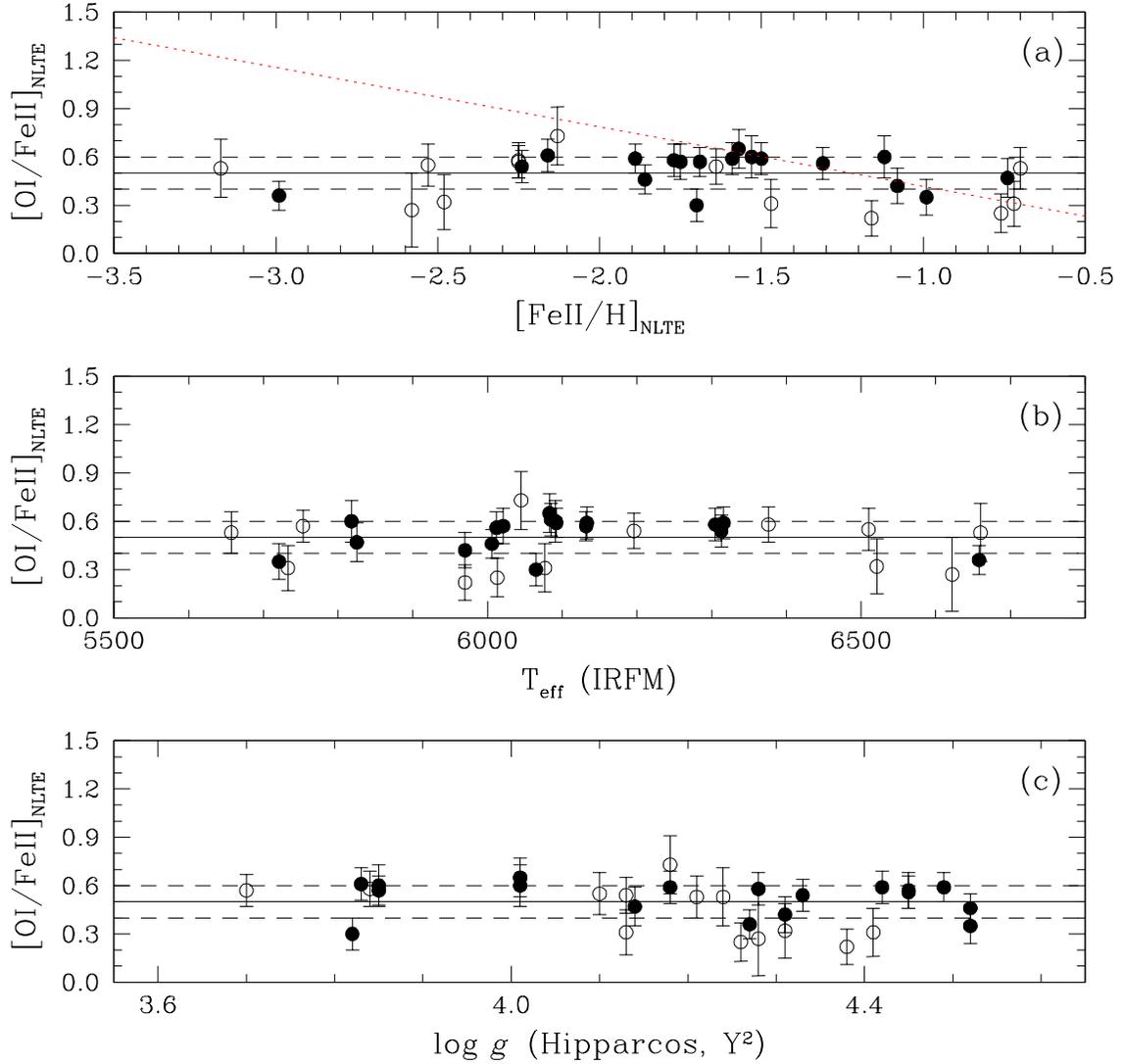}
\caption{Oxygen-to-iron ratios vs. iron abundance (a), 
\teff (b) and log $g$ (c). 
Filled and open circles represent stars with uncertainties lower
and higher than $\Delta$$E_{B-V}$ = 0.01 mag, respectively.
The mean [O/Fe] = +0.5 is shown by solid lines, and 
$\pm$0.1 dex uncertainties are shown by dashed lines.
A linear relation between [O/Fe] and [Fe/H] previously found in other 
studies of the \ion{O}{1} triplet (Mishenina et al. 2000; 
Israelian et al. 1998, 2001; Boesgaard et al. 1999) is shown
by a dotted line with a slope = $-$0.35 (top panel)}
\label{ofe}
\end{figure}

\begin{figure}
\epsscale{}
\plotone{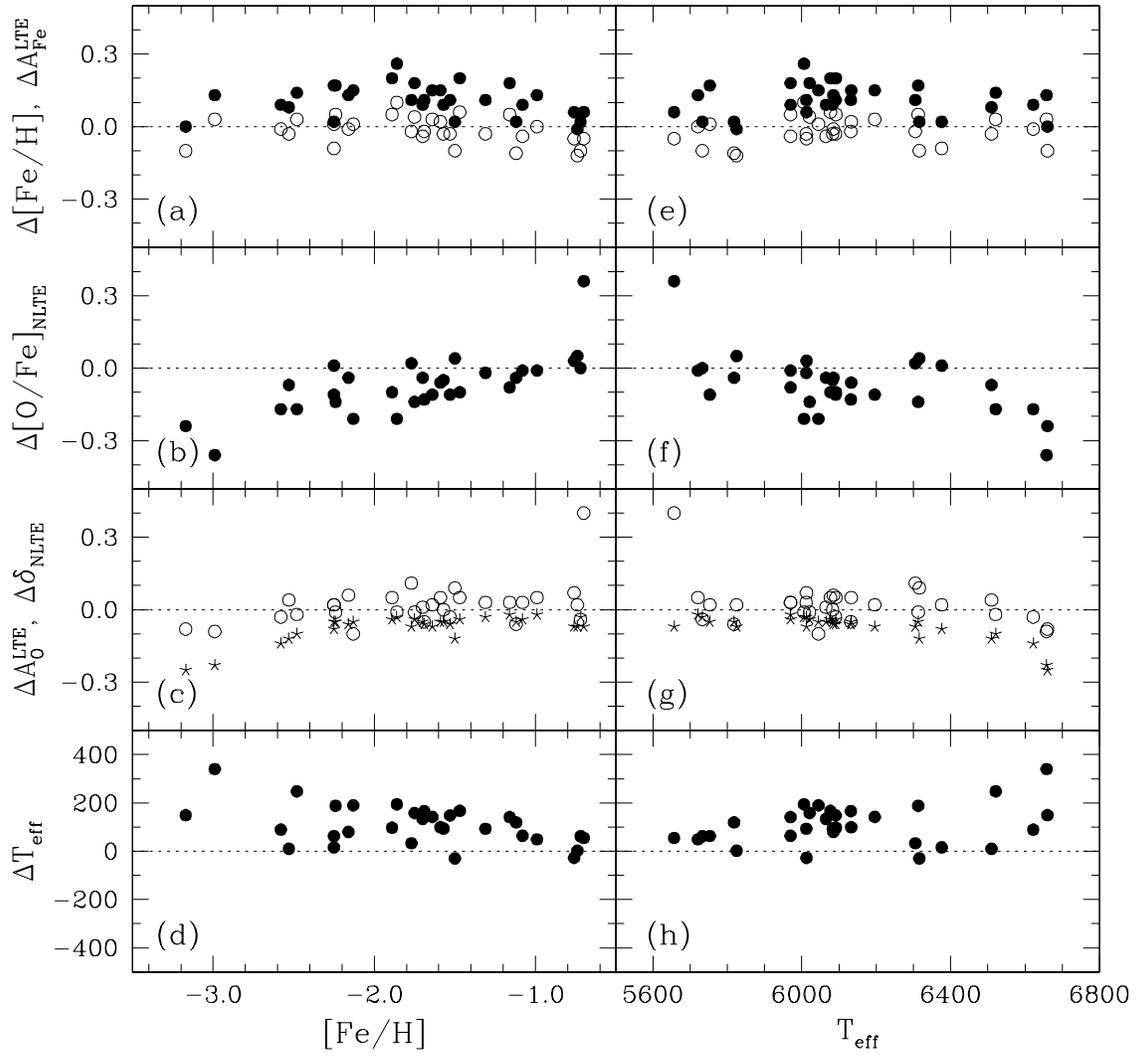}
\caption{Differences between the present work and Ake04
are shown in the left and right panels as a function 
of [Fe/H] and \tsin, respectively:
$\Delta$[Fe/H] (filled circles); $\Delta$A$_{Fe}^{\rm LTE}$ (open circles);
$\Delta$[O/Fe]$_{\rm NLTE}$ (filled circles);
$\Delta$A$_{O}^{\rm LTE}$ (open circles); $\Delta \delta_{\rm NLTE}$ (stars);
$\Delta$\teff (filled circles).}
\label{ake04}
\end{figure}

\begin{figure}
\epsscale{}
\plotone{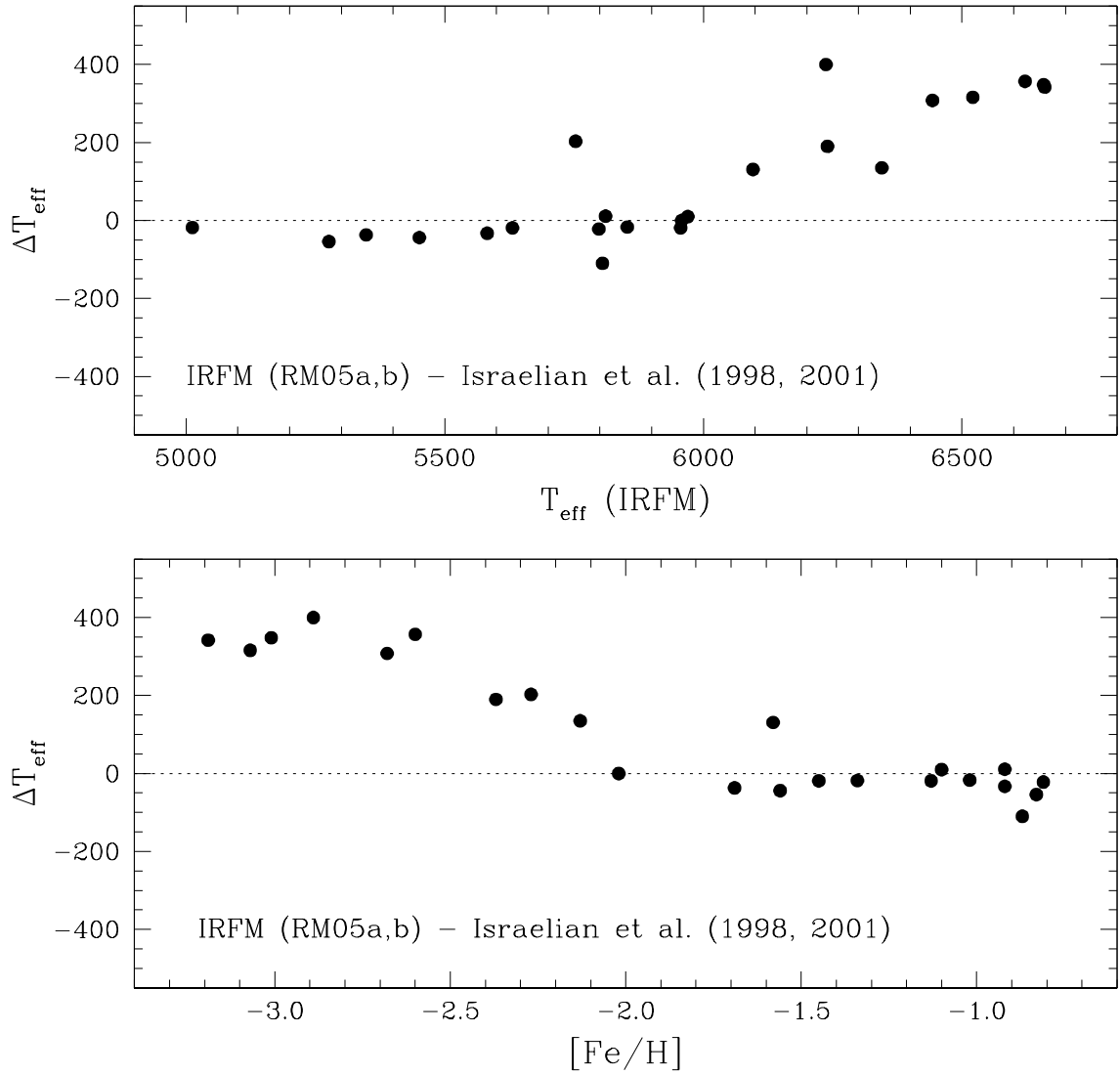}
\caption{Temperatures obtained by us employing the IRFM \teff scale
of RM05a,b minus the temperatures
from Israelian et al. (1998, 2001) as a function of
\teff (upper panel) and [Fe/H] (lower panel).}
\label{ofe}
\end{figure}

\begin{figure}
\epsscale{}
\plotone{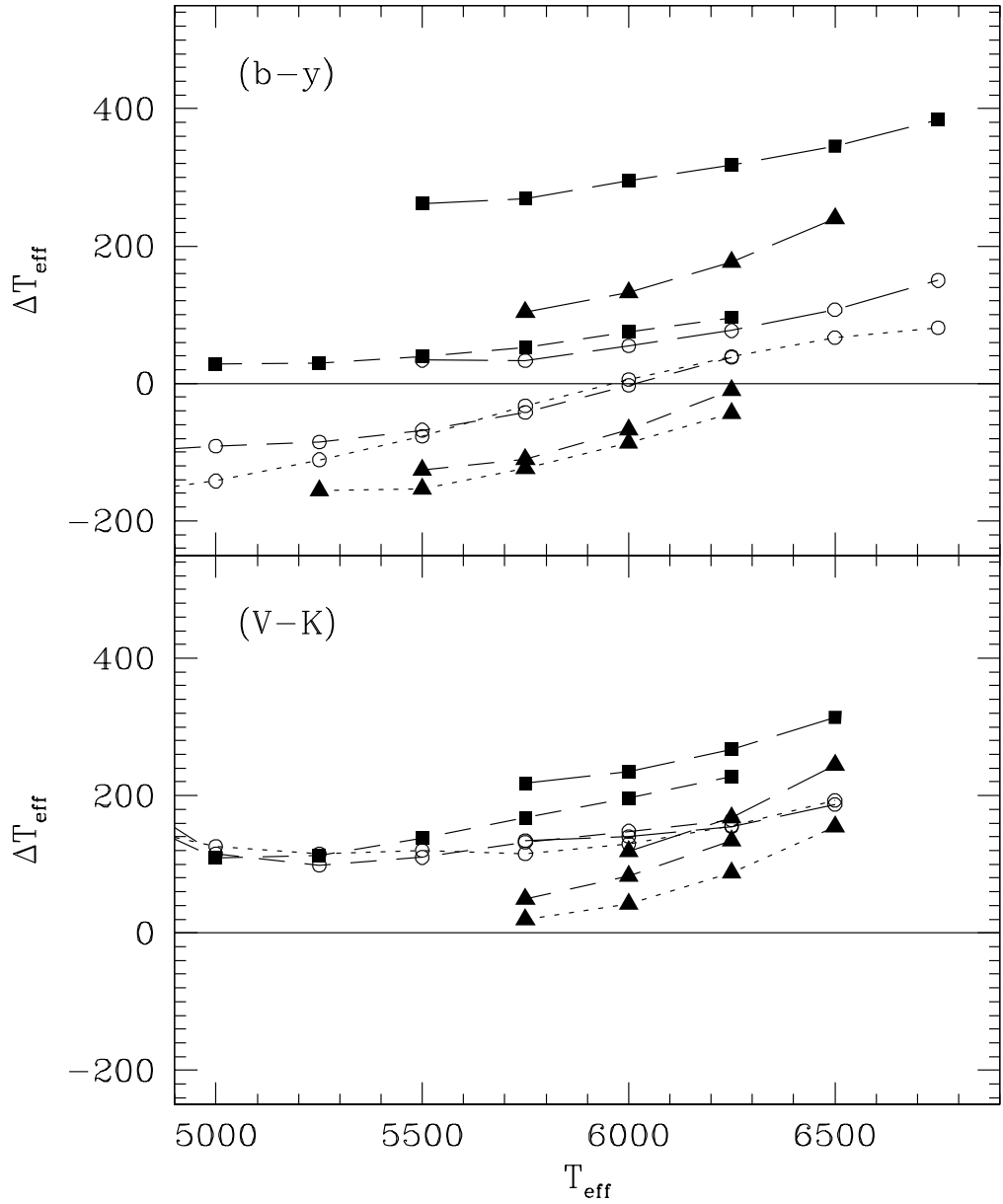}
\caption{Upper panel: difference between the (b-y) temperature scales by RM05b
and: Alonso et al. (1996, circles), Carney (1983, squares), and King (1993, triangles); 
at [Fe/H]= $-$1 (dotted lines), [Fe/H] = $-$2 (dashed lines), and [Fe/H] = $-$3 (long dashed
lines). The solid line is at $\Delta$ \teff = 0. Lower panel: same but for (V-K).
}
\label{byVK}
\end{figure}

\begin{figure}
\epsscale{}
\plotone{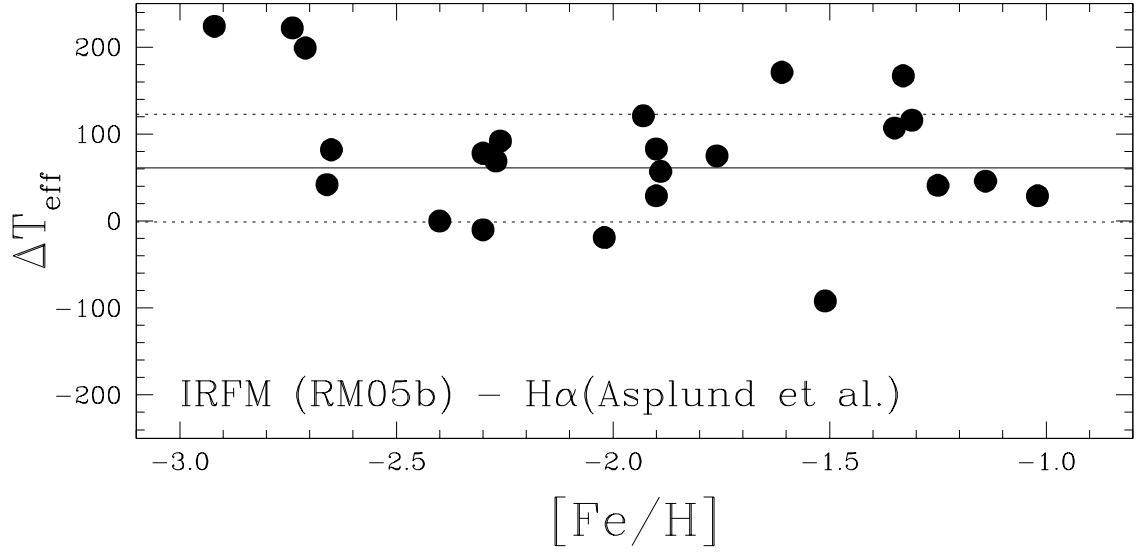}
\caption{Temperatures employing the IRFM \teff scale by
RM05b minus the H$\alpha$ temperatures determined
by Asplund et al. (2005). Excluding the three stars with
[Fe/H] $< -$2.7, the zero point shift is 61$\pm$62 K,
which is represented by a solid line. Dotted lines
represent the scatter ($\sigma$ = 62 K). 
}
\label{ofeohuv}
\end{figure}

\begin{figure}
\epsscale{}
\plotone{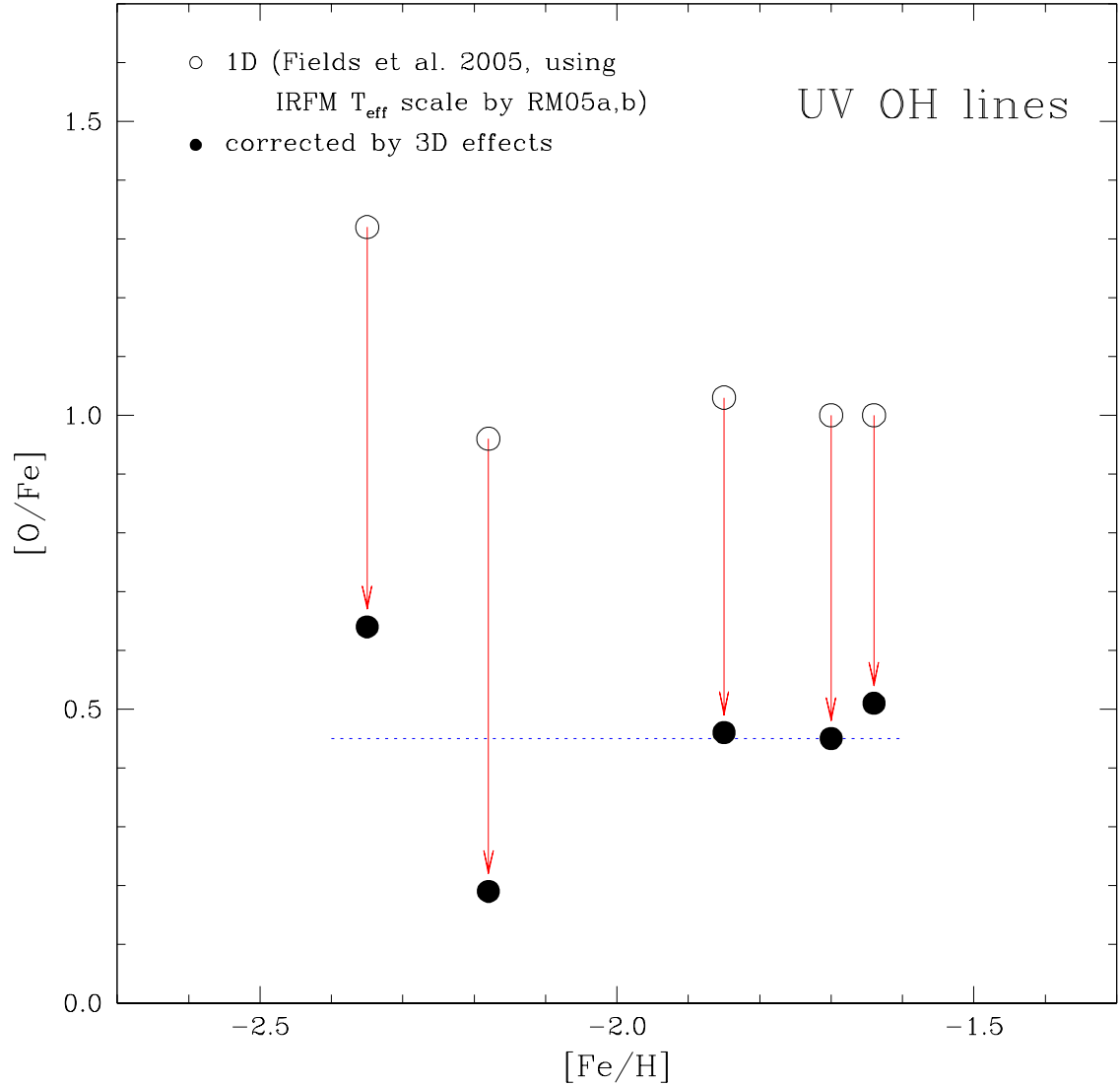}
\caption{[O/Fe] (from UV OH lines) vs. [Fe/H] employing
the new \teff scale by RM05a,b. 
Open circles represent the [O/Fe] ratios obtained 
in the 1D analysis by Fields et al. (2005);
filled circles represent the [O/Fe] after 3D effects (arrows) 
are taken into account (from 3D corrections computed by 
Asplund \& Garc\'{\i}a P\'erez 2001). The dotted line
represents the mean 3D-corrected [O/Fe] = +0.45 dex.
}
\label{ofeohuv}
\end{figure}

\begin{figure}
\epsscale{}
\plotone{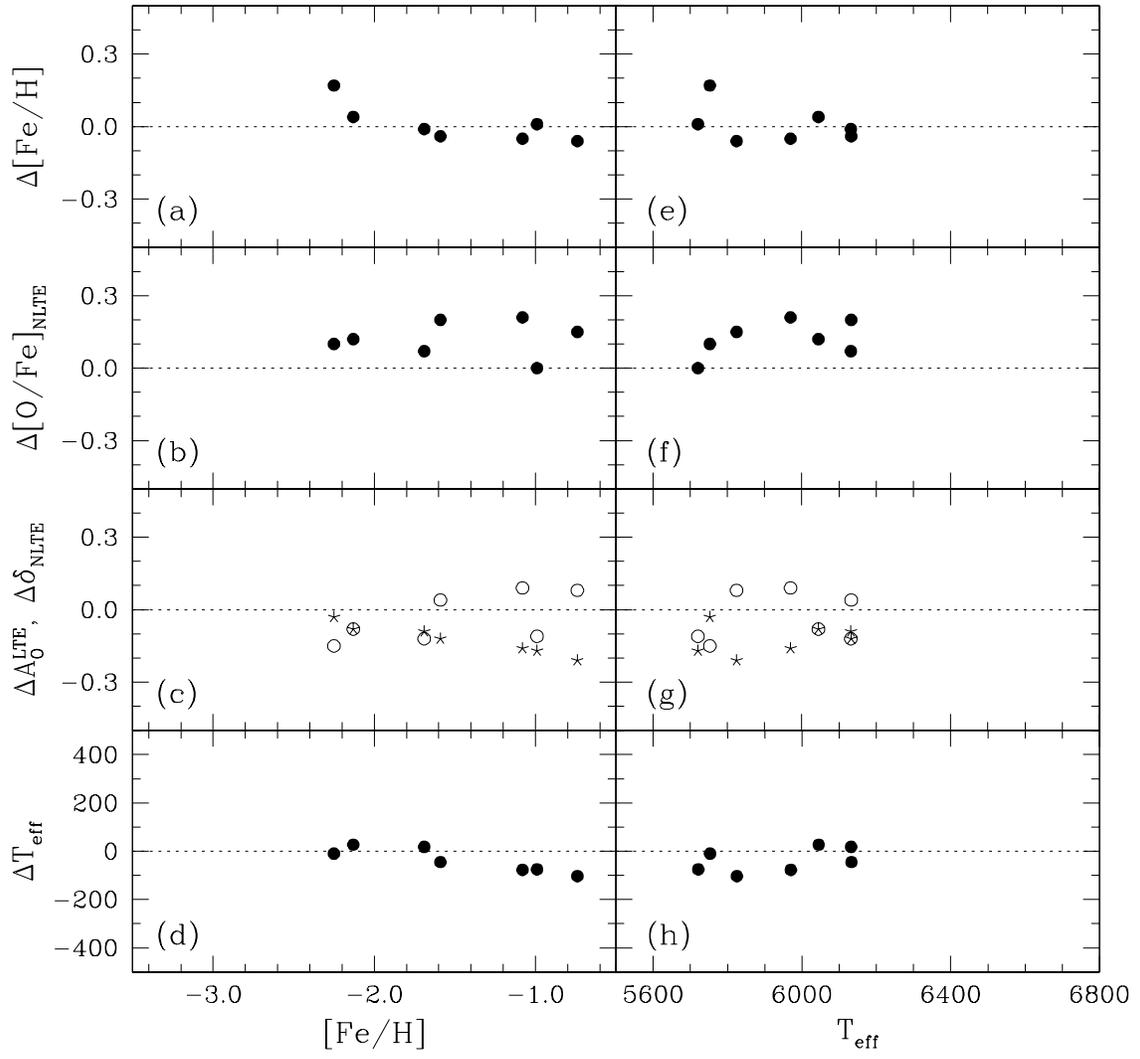}
\caption{Differences between the present work and Carretta et al. (2000)
are shown in the left and right panels as a function 
of [Fe/H] and \tsin, respectively:
$\Delta$[Fe/H] (filled circles); 
$\Delta$[O/Fe]$_{\rm NLTE}$ (filled circles);
$\Delta$A$_{O}^{\rm LTE}$ (open circles); $\Delta \delta_{\rm NLTE}$ (stars);
$\Delta$\teff (filled circles).}
\label{carretta}
\end{figure}

\begin{figure}
\epsscale{}
\plotone{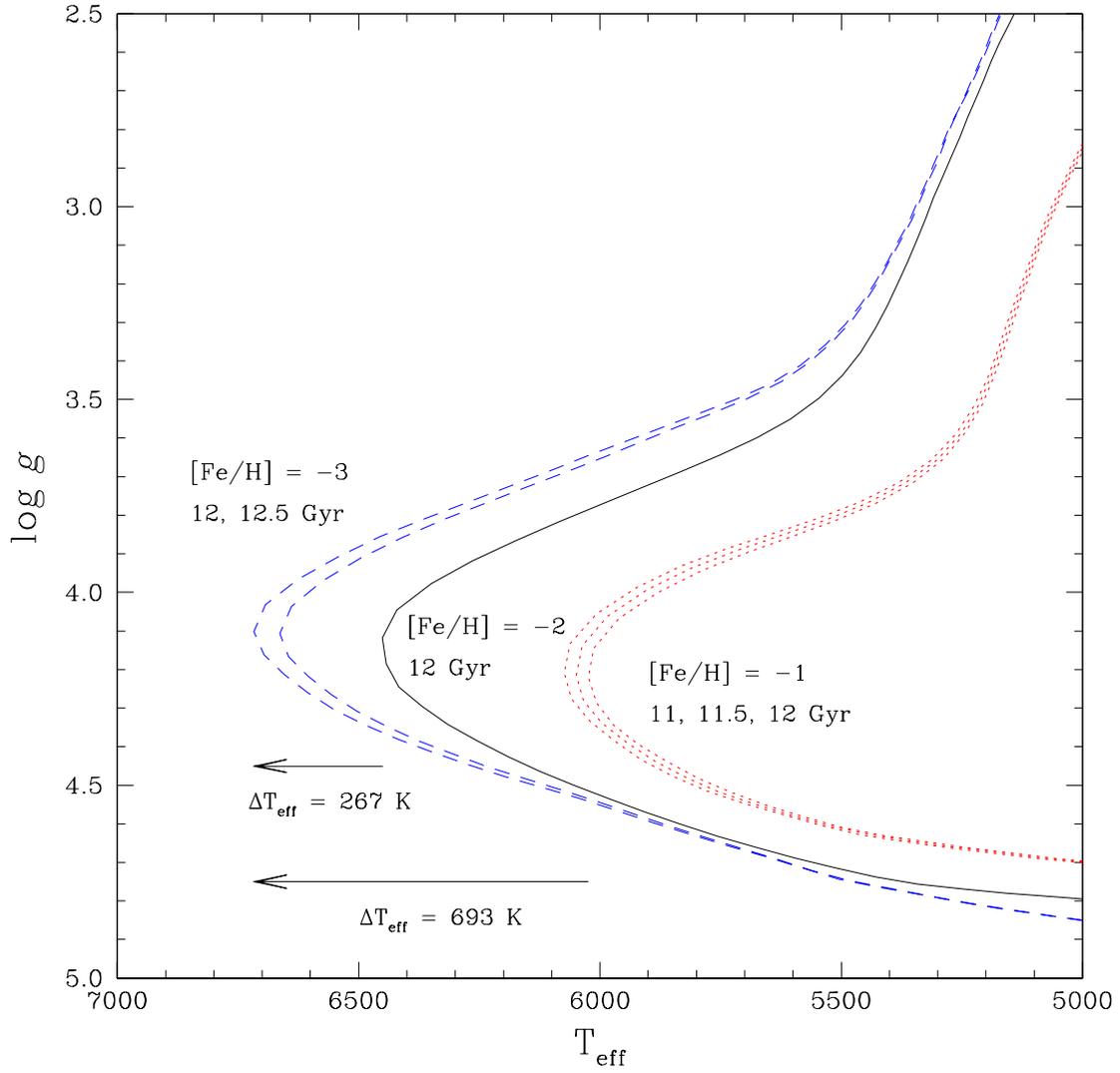}
\caption{Y$^2$ isochrones for [Fe/H] = -1 (dotted lines), -2 (solid line)
and -3 (dashed lines). The arrows show the maximum increase in
\teff of the turn-off from [Fe/H] = -1 to -3 (693 K), and from [Fe/H] = -2 to -3 (267 K).}
\label{isoc}
\end{figure}

\begin{figure}
\epsscale{}
\plotone{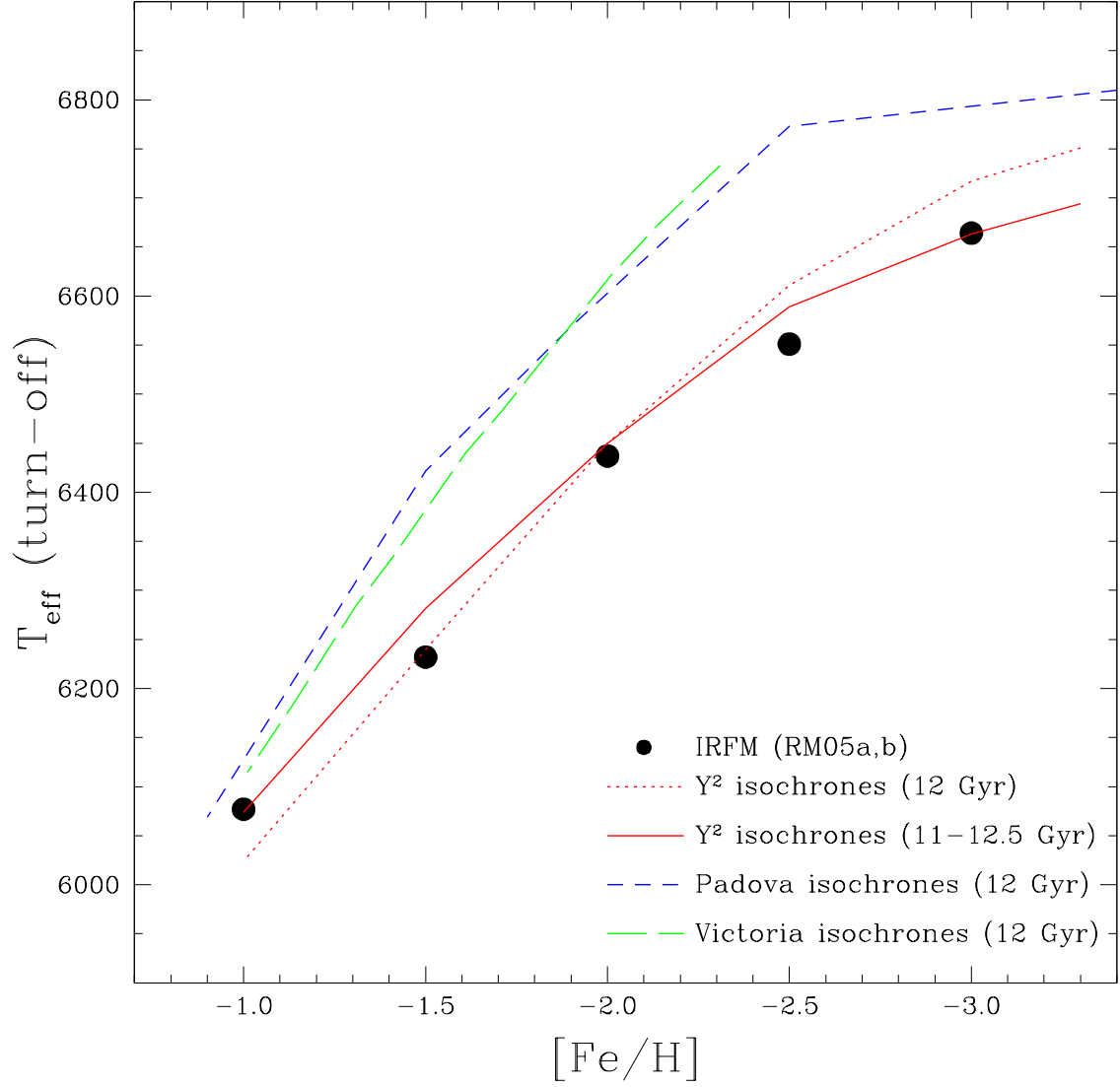}
\caption{Filled circles: Temperatures of turn-off stars found in this work and
in Mel\'endez \& Ram\'{\i}rez (2004), employing the IRFM
\teff scale by RM05a,b. Dotted line: predictions of 
Y$^2$ isochrones adopting an age of 12 Gyr. Solid line: predictions
of Y$^2$ isochrones for ages increasing from 11 Gyr ([Fe/H] = -1)
to 12.5 Gyr ([Fe/H] = -3.3). Short and long dashed lines: 12 Gyr turn-off \teff 
from Padova and Victoria isochrones, respectively.
}
\label{isoc}
\end{figure}

\end{document}